\documentclass[11pt]{article}
\usepackage{latexsym,amssymb}
\newcommand{\ft}[2]{{\textstyle\frac{#1}{#2}}}
\def\tilde{\widetilde}

\def\1bar{1\hskip -.275cm -}
\def\2bar{2\hskip -.275cm -}
\def\3bar{3\hskip -.275cm -}

\newsavebox{\uuunit}
\sbox{\uuunit}
    {\setlength{\unitlength}{0.825em}
     \begin{picture}(0.6,0.7)
        \thinlines
        \put(0,0){\line(1,0){0.5}}
        \put(0.15,0){\line(0,1){0.7}}
        \put(0.35,0){\line(0,1){0.8}}
       \multiput(0.3,0.8)(-0.04,-0.02){10}{\rule{0.5pt}{0.5pt}}
     \end {picture}}

\makeatletter \@addtoreset{equation}{section} \makeatother

\def\bfone{\relax{\rm 1\kern-.35em 1}}

\def\bfone{\relax{\rm 1\kern-.35em 1}}
 
\begin{document}
\begin{titlepage}
\begin{flushright}
NORDITA-2001/11 HE\\
SPIN-2001/15
\end{flushright}
\vskip 1.5cm
\begin{center}
{\LARGE \bf   Supersymmetric 3-branes on smooth}\\
{\LARGE \bf  ALE manifolds with flux $^\dagger$}\\ \vfill {\large
M. Bertolini$^1$, V. L. Campos$^2$,
G. Ferretti$^2$, \\ P. Fr\'e$^3$, P. Salomonson$^2$,  M. Trigiante$^4$} \\
\vfill { \sl $^1$  NORDITA, Blegdamsvej, 17
2100 Copenhagen \O, Denmark\\
\vskip 0.3cm
$^2$ Institute for Theoretical Physics - G\"oteborg University and \\
Chalmers University of Technology, 412 96 G\"oteborg, Sweden
\vskip 0.3cm
$^3$ Dipartimento di Fisica Teorica, Universit\'a di Torino, $\&$ INFN -
Sezione di Torino\\
via P. Giuria 1, I-10125 Torino, Italy \\
\vskip 0.3cm $^4$ Spinoza Institute, Leuvenlaan 4, Utrecht, The
Netherlands }
\end{center}
\vfill
\begin{abstract}
{We construct a new family of classical BPS solutions of type IIB
supergravity describing 3-branes transverse to a 6-dimensional space
with topology  $\mathbb{R}^{2}\times$ALE. They are characterized by a
non-trivial flux of the supergravity 2-forms through the homology $2$-cycles of
a generic smooth ALE manifold.
Our solutions have two Killing spinors and thus preserve $\mathcal{N}=2$
supersymmetry. They
are expressed in terms of a quasi harmonic function $H$
(the ``warp factor''), whose properties we study in the case
of the simplest ALE,
namely the Eguchi-Hanson manifold. The equation for $H$
is identified as an instance of the confluent Heun equation. We
write explicit power series solutions and solve the
recurrence relation for the coefficients, discussing also the
relevant asymptotic expansions. While, as in all such
$\mathcal{N}=2$ solutions, supergravity breaks down near the brane,
the smoothing out of the vacuum geometry has the effect that the warp factor
is regular in a region near the cycle. We interpret the behavior of the warp
factor as describing a three-brane charge ``smeared'' over the cycle and
consider the asymptotic form of the geometry in that region, showing that
conformal invariance is broken even when the complex type IIB 3-form field
strength is assumed to vanish. We conclude with a discussion of the basic
features of the gauge theory dual.

}
\end{abstract}
\vspace{2mm} \vfill \hrule width 3.cm {\footnotesize $^ \dagger $
This work is supported in part by the European Union RTN contracts
HPRN-CT-2000-00122 and HPRN-CT-2000-00131.}
\end{titlepage}

\tableofcontents

\section{Introduction}
\label{intro}
After the seminal paper by Maldacena \cite{mal},
many efforts have been devoted to extend the gauge/gravity
correspondence to less supersymmetric and non-conformal cases. In
this context considerable attention was recently directed to
the study of fractional branes
\cite{Klebanov:2000rd}-\cite{marco}. These are the natural
elementary D-branes occurring whenever string theory is reduced on
a (not necessarily compact) orbifold \cite{Anselmi:1994sm},
\cite{Douglas:1996sw}-\cite{Diaconescu:1998br}.
Because of their nature they cannot
move away from the orbifold apex and thus the dual gauge
theory on their world-volume lacks the relevant moduli fields.
Generically, this leads to both reduced supersymmetry and
non-vanishing $\beta$-functions. Most interesting are the fractional
D3-branes, namely the case when the world-volume theory is
four-dimensional. In this respect the two most appealing
situations are provided by the ${\cal N}=1$ case emerging from
singular limits of CY spaces  and the $\mathcal{N}=2$ case arising
from the singular limit of ALE spaces. Much work was devoted to
both.

A common feature of many supergravity solutions representing
non-conformal situations is the presence of naked singularities of
repulson type \cite{kal}. These correspond to IR singularities at
the gauge theory level and one expects that they should be resolved or
explained by some stringy effect. Although a general recipe
does not seem to exist, progress in understanding such an issue was
made both for the ${\cal N}=1$ and the $\mathcal{N}=2$ case.
\par
In constructing supergravity solutions corresponding to fractional
branes one proceeds in two steps. Firstly, one defines the vacuum
solution as a space-time obtained by quotiening ten-dimensional
Minkowski space-time modulo a discrete group with all other fields
set to their constant background values (typically, the latter are
zero with the exception of the dilaton, which sets the
string coupling, and the NS-NS 2-form $B$, which determines whether
the conformal field theory is well defined or not \cite{bflux}).
Secondly, one places the branes at the orbifold/conifold apex,
which has the effect of turning on the relevant gauge fields and
of warping the metric in a way that makes the curvature non-zero
even away from the singularity. As already stressed, these
solutions generally display singularities (see \cite{KLEBA2} for
the ${\cal N}=1$ case and \cite{Bertolini:2000dk} for the ${\cal
N}=2$ case).

Once confronted with superstring theory on such a space-time, it
is a natural question to ask whether this space can be thought of
as a singular limit of a smooth geometry. This can always be
achieved when no branes are present. If the smoothing out can be
carried out even in the presence of branes, and if the warp factor
is such that the curvature remains small everywhere and the
singularities in the solution are removed, then one can claim that
the supergravity solution provides a dual description of the gauge
theory. This is what was accomplished, for instance, in the ${\cal
N} =1$ {\em deformed} conifold solution of Klebanov and Strassler
\cite{KLEBA3}, where the dynamics of the brane  deforms
the conifold at its apex, yielding a singularity free solution.
\par
What is somewhat frustrating is that, contrary to what one would
naively expect, the ${\cal N}=1$ case turned out to be more
amenable to a smooth supergravity analysis than the $\mathcal{N}=2$
case. Actually, for
the $\mathcal{N}=2$ case, a completely different phenomenon has been
shown to occur \cite{Bertolini:2000dk}-\cite{Petrini:2001fk}: the
singular region is excised by the enhan\c{c}on mechanism
\cite{enhanc}, which corresponds to a locus where extra massless
degrees of freedom come into play, thus completely changing the
geometry in the interior. From the gauge theory point of view the
enhan\c{c}on is the scale at which the gauge coupling diverges and
where instantonic effects become relevant.
\par
Despite this specific feature of the $\mathcal{N}=2$ case, one could
anyhow try to see if better behaved solutions can arise when
considering, analogously to what has been done in the conifold
case in \cite{KLEBA3} and \cite{TSEYTLIN1}, the blow-up of the
singular space, namely the smooth ALE manifold. Then one could try to
see if there is any relation between the enhan\c{c}on and the
geometric moduli of the better behaved geometry, try to
disentangle the singular properties of the solution, and
eventually investigate their meaning in the context of
the dual gauge theory.
\par
Starting from the singular space $\mathbb{C}^2/\Gamma$, where $
\Gamma$ is a discrete subgroup of $SU(2)$, one should inquire
whether one can soften  the vacuum geometry to a
smooth ALE space. It is well known that this {\it cannot} be done
in a supersymmetric way if fractional D3-branes are present and if
the string theory is really at the orbifold point, that is, with a
non zero background $B$-field threaded through the vanishing cycle
of the orbifold. There are many arguments for such a ``no-go''
theorem. Probably the most intuitive one involves going to the
T-dual type IIA picture, \cite{das}. There one finds two NS5-branes
with a stack of D4-branes (dual to the fractional D3-branes)
stretched between them. The blowing up of the orbifold corresponds
to moving the NS5-branes in the direction transverse to the
D4-branes, and this involves stretching the D4-branes even further,
thus increasing the energy of the system and hence breaking
supersymmetry\footnote{We thank R. Russo for a discussion on this
point.}.
\par
Yet the above argument  does not rule out the existence of supersymmetric
{\em bulk} solutions of type IIB supergravity that smooth out the
quotient. Such solutions, which  eventually have to be interpreted as
suitable bound states of some D-brane state, are precisely those
that we find in the present paper.
\par
In particular, we find ${\cal N} =2$ BPS solutions
corresponding to 3-brane states where the transverse space is given
by $\mathbb{R}^2\times {\rm ALE}$ and where there is a non
zero value of the complex type IIB supergravity 3-form in the transverse
directions (note that this is the distinctive
feature of fractional D-branes in the singular orbifold theories).
As already noticed, solutions of this type with ${\cal N} = 1$
supersymmetry have appeared in the recent literature but they were
mostly concerned with the case where the six-dimensional
transverse space was a cone. This had the simplifying feature of
making the warp factor depend only on one (radial) coordinate.
This feature cannot be preserved if one imposes ${\cal N} =
2$ supersymmetry since the required $\mathrm{SU(2)}$ holonomy forces the
geometry of the transverse space to be of the factorized form
described above. The price to pay is that the warp factor is now
dependent on (at least) two variables and it is
governed by a complicated partial differential equation. In fact
the whole set of IIB equations can be reduced to the solution of
such an equation for the warp factor, whose source is essentially
dictated by supersymmetry up to an arbitrary analytic
function $\gamma(z)$.
\par
Supergravity alone is not sufficient to determine $\gamma(z)$ or the
boundary conditions.
This arbitrariness implies that our solution describes various
deformation or various vacua of ${\cal N}=2$ theories. The case
$\gamma(z)= \hbox{const.}$ describes a vanishing three-form flux and
corresponds to the well known conformal
${\cal N}=2$ theory with product gauge group $\mathrm{U(N)} \times \mathrm{U(N)}$,
hyper-multiplets in the bi--fundamental representation and Fayet-Iliopoulos terms describing
the ALE moduli. The new ingredient in our construction is the following. We show how at
the supergravity level a three-form flux
can be turned on consistently with supersymmetry.  Some consequences of
this for the gauge dual theory are analyzed in the concluding section
yet a complete analysis of such an issue involves a systematic study of possible
supersymmetric world-volume actions of branes on arbitrary curved
backgrounds (in particular ALE manifolds) which is work in progress and the subject of a
forthcoming publication.
\par
The paper is organized as follows.
In section \ref{bosact2b} we briefly recall the structure of the
equations of motion of type IIB supergravity, fixing our
conventions for the fields. In section \ref{3}, after
studying the general set-up for arbitrary ALE spaces, we give the
structure of our 3-brane solution whose distinctive feature is the
localized 3-form flux along any exceptional compact 2-cycle of the
ALE space. Finally, applying a criterion which is analogous to
that discussed in \cite{GRANA}-\cite{CVETIC2}, (see also \cite{ke})
we give the proof
of its supersymmetry (in this respect, our solutions fall in the
so called type B class \cite{GRANA2}, characterized by the 3-form to
be imaginary self-dual). In order to consider an explicit example, in
section \ref{Ehcase} we specialize to the simplest such space, the
Eguchi-Hanson space, and study in detail the properties of the
warp factor. Finally, in section \ref{conc} we summarize our results,
discuss the relevant physical properties of our solutions, discuss
some basic aspects of their gauge theory dual
and conclude with some comments on open issues.

\section{Bosonic action and field equations of type IIB supergravity}
\label{bosact2b}
As it is well known, type IIB supergravity does not have any conventional
supersymmetric action. However, as it happens for all on-shell
supergravity theories, the complete set of field equations can be
obtained as consistency conditions from the closure of the supersymmetry
transformation algebra \cite{Schwarz:1983qr}. In the case of type
IIB supergravity,  one was also able \cite{igorleo} to obtain a
complete, manifestly $\mathrm{SU(1,1)}$-covariant formulation of the theory
based on the rheonomic approach to supergravity theory
\cite{castdauriafre}.
\par
The bosonic part of the equations can be formally obtained through
variation of the following action \footnote{Note that our $R$ is
equal to $- \ft 1 2 R^{old}$, $R^{old}$ being the normalization of
the scalar curvature usually adopted in General Relativity
textbooks. The difference arises because in the traditional
literature the Riemann tensor is not defined as the components of
the curvature $2$-form $R^{ab}$ rather as $-2$ times such
components.}:
\[
S_{\rm IIB} = \frac{1}{2 \kappa^2} \Bigg\{ \int d^{10} x~ \left[
-2  \sqrt{-\det g}~ R \right] - \frac{1}{2} \int \Big[ d \varphi
\wedge \star d \varphi
 \,+\, {\rm e}^{- \varphi} F_{[3]}^{NS}  \wedge \star F_{[3]}^{NS}\,+\, {\rm e}^{2
 \varphi}\, F_{[1]}^{RR} \wedge \star F_{[1]}^{RR}
\]
\begin{equation}
 + \,\,{\rm e}^{\varphi} \,{F}_{[3]}^{RR} \wedge \star
 {F}_{[3]}^{RR} \,+\, \frac{1}{2}\, {F}_{[5]}^{RR}
 \wedge \star {F}_{[5]}^{RR}  \, -\,  C_{[4]} \wedge
 F_{[3]}^{NS}
 \wedge F_{[3]}^{RR} \Big] \Bigg\}
\label{bulkaction}
\end{equation}
where:
\begin{eqnarray}
F^{RR}_{[1]} & = & dC_{[0]} \nonumber\\
F^{NS}_{[3]} & = & dB_{[2]} \nonumber\\
F^{RR}_{[3]}& = & dC_{[2]} -  \, C_{[0]} \,
dB_{[2]}\nonumber\\
F^{RR}_{[5]}& = & dC_{[4]}- \ft 12 \left( B_{[2]} \wedge d C_{[2]}
-  C_{[2]} \wedge d B_{[2]}\right) \label{bosecurve}
\end{eqnarray}
It is important to stress though that the action
(\ref{bulkaction}) is to be considered only a book keeping device
since the $4$-form $C_{[4]}$ is not free, its field strength
$F_{[5]}^{RR}$  being subject to the on-shell self-duality
constraint:
\begin{equation}
F_{[5]}^{RR} = \star F_{[5]}^{RR} \label{selfonshell}
\end{equation}
>From the above action the corresponding equations of motion can be
obtained:
\begin{eqnarray}
d \star d \varphi - e^{2\varphi} \, F^{RR}_{[1]} \wedge \star
F^{RR}_{[1]} & = & -\ft 1 2 \, \left( e^{-\varphi} F^{NS}_{[3]}
\wedge \star  F^{NS}_{[3]}-
  e^{\varphi} F^{RR}_{[3]} \wedge \star  F^{RR}_{[3]}\right) \label{NSscalapr}\\
  d\left( e^{2\varphi} \star F^{RR}_{[1]}\right)  & = & - e^{\varphi} \, F^{NS}_{[3]} \wedge \star  F^{RR}_{[3]}
\label{RRscalapr}\\
d\left( e^{-\varphi} \, \star F_{[3]}^{NS}\right) + e^\varphi \,
F^{RR}_{[1]} \wedge \star F^{RR}_{[3]}
  & = &  - F_{[3]}^{RR} \wedge F^{RR}_{[5]}
\label{3formNS}\\
d\left( e^\varphi \star F_{[3]}^{RR } \right) & = & -F_{[5]}^{RR}
\, \wedge F_{[3]}^{NS}
\label{3formRR}\\
d\star F^{RR}_{[5]} & = & -
  F^{NS}_{[3]} \, \wedge \, F^{RR}_{[3]}
\label{f5RR}\\
-\,2 \,R_{{MN}}&=& \frac{1}{2}\partial_{{M}}\varphi
\partial_{{N}}\varphi+\frac{e^{2\varphi}}{2}
\partial_{{M}} C_{[0]} \partial_{{N}}
C_{[0]}+150
 {F}_{[5]{M}\cdot\cdot\cdot\cdot}
{F}_{[5]{N}}^{\phantom{{M}}\cdot\cdot\cdot\cdot}
\nonumber\\
& &+ 9 \left( e^{-\varphi}F_{[3]{M}\cdot\cdot}^{NS}\,
F_{[3]{N}}^{{NS}\phantom{{M}}\cdot\cdot} +e^{\varphi}{
F}_{[3]{M}\cdot\cdot}^{RR}
{ F}_{[3]{N}}^{RR\phantom{{M}}\cdot\cdot}\right)\nonumber\\
& & -\frac{3}{4}\,
g_{{MN}}\,\left(e^{-\varphi}F_{[3]\cdot\cdot\cdot}^{NS}
F_{[3]}^{NS\cdot\cdot\cdot}+e^{\varphi}{{F}}_{[3]\cdot\cdot\cdot}^{RR}{
F}^{RR\cdot\cdot\cdot}_{[3]}\right) \label{einsteinequa}
\end{eqnarray}
It is not difficult to show, upon suitable identification of the
massless superstring fields, that this is the correct set of
equations which can be consistently obtained from the manifestly
$\mathrm{SU(1,1)}$ covariant formulation of type IIB supergravity
\cite{igorleo}.

\section{3-brane solution with ALE flux}
\label{3}
In this section we provide the BPS solution corresponding to a
3-brane transverse to a smooth ALE space, namely we construct type IIB
supergravity solutions describing 3 branes on a vacuum
$\mathbb{R}^{1,3} \times \mathbb{R}^{2}\times \mathrm{ALE}$. This will be
achieved without an analysis of the specific form of the
world-volume action of the brane, i.e. of the source term. Our
physical assumption will just be that, together with the usual RR
5-form flux, the 3-brane solution has a non-trivial flux of the
supergravity 2-form potentials along (one of) the compact two cycle(s)
of the blown-up orbifold (this translates into a non-vanishing value of the complex 3-form
field strength).

\subsection{Solution of the bosonic field equations}
\label{sol3flux}
We separate the ten coordinates of space-time into the following subsets:
\begin{equation}
 x^M = \left \{ \begin{array}{rcll}
&x^\mu \quad \mu =0,1,2,3& \mbox{coordinates of the 3-brane world volume}   \\
&z  \;=\;  x^4 + {\rm i} x^5 & \mbox{complex coordinate of
$\mathbb{R}^2 \sim \mathbb{C}$} \null \\
&y^\tau \quad \tau=6,7,8,9 & \mbox{real coordinates of the ALE
4-space }   \
\end{array} \right.
\label{coordisplit}
\end{equation}
and we make the following ansatz for the metric\footnote{As
explained in appendix A, the conventions for the gamma matrices
and the spinors are set with a mostly minus metric $d\tau^2$. In the
discussion of the solution, however, we use $ds^2 = -d\tau^2$ for
convenience. We hope this does not cause any confusion.}:
\begin{eqnarray}
ds^2&=&H^{-\frac{1}{2}}\left (-\eta_{\mu\nu}dx^\mu\,dx^\nu \right
)+H^{\frac{1}{2}}dzd\bar{z}+H^{\frac{1}{2}}
ds^2_{ALE}\label{ansazzo}\\
H&=&H(z,\bar{z},y)\nonumber\\
ds^2_{ALE}&=&g^{ALE}_{\tau\sigma} dy^\tau\,dy^\sigma\nonumber\\
{\rm det}(g)&=&H{\rm det}(g^{ALE})\nonumber\\
\eta_{\mu\nu}&=&{\rm diag}(+,-,-,-)
\end{eqnarray}
where $g^{ALE}_{\tau\sigma}$ is the metric of any ALE space and we
denote the six-manifold spanned by $z$, $\bar z$ and $y^\tau$ by
\begin{equation}
  M_6 = \mathbb{R}^2 \times \mathrm{ALE}
\label{m6defi}
\end{equation}
In terms of vielbein the ansatz (\ref{ansazzo})  corresponds to
\begin{equation}
  V^{A}= \cases {V^a = H^{-1/4} (z, \bar z, y) \, dx^a \cr
  V^i = H^{1/4} (z, \bar z, y) e^i \cr}
\label{splittoviel}
\end{equation}
where   $e^i$ are the vielbein $1$-forms of the manifold $M_6$.
The structure equations of the latter are\footnote{The hats over
the spin connection and the Riemann tensor denote quantities
computed without the warp factor.}:
\begin{eqnarray}
0& = &  d \, e^i  - \widehat{\omega}^{ij} \, \wedge \, e^k \, \eta_{jk}\nonumber\\
\widehat{R}^{ij} & = & d \widehat{\omega}^{ij} -
\widehat{\omega}^{ik}\, \wedge \, \widehat{\omega}^{\ell j} \,
\eta_{k\ell} = \widehat{R}^{ij} _{\phantom{ij}\ell m } \, e^\ell
\,\wedge \, e^m \label{structeque}
\end{eqnarray}
The relevant property  of the $M_6$ metric that we use in solving
the Einstein's equations is that it is Ricci flat:
\begin{equation}
 \widehat{R}^{im}_{\phantom{ij}\ell m } = 0 \label{ricciflatto}
\end{equation}
What we need in order to derive our solution and discuss its
supersymmetry properties is the explicit form of the spin
connection for the full $10$-dimensional metric (\ref{ansazzo})
and  the corresponding Ricci tensor. From the torsion equation one
can uniquely determine the solution:
\begin{eqnarray}
\omega^{ab} & = & 0 \nonumber\\
\omega^{ai} & = &  \ft 1 4 \, H^{-3/2} \, dx^a  \eta^{ik} \,
\partial_k \, H
\nonumber\\
\omega^{ij} & = & \widehat{\omega}^{ij} + \Delta \omega^{ij} \quad
; \quad \Delta \omega^{ij} =- \ft 1 2 \, H^{-1}  \, e^{[i}  \,
\eta^{j]k} \, \partial_k H \label{spinconnect}
\end{eqnarray}
Inserting this result into the definition of the curvature
$2$-form we obtain:
\begin{eqnarray}
R^{a}_{b} & = & -  \frac{1}{8}\, \left [ H^{-3/2} \square_6  \, H -
H^{-5/2} \,
\partial_k H\partial^k H \right] \, \delta^a_b \nonumber\\
R^{a}_{i} & = & 0\nonumber\\
R_j^i   &=& \frac{1}{8}H^{-3/2} \Box_6 H\delta_j^i
             - \frac{1}{8} H^{-5/2}\partial_k H\partial^k H\delta_j^i
             +\frac{1}{4} H^{-5/2} \partial_j H\partial^i H
\label{riccius}
\end{eqnarray}
where
\begin{equation}
\square_6 \equiv \square_{\mathbb{R}^2} \, + \, \square_{\mathrm{ALE}}
\label{laplacious}
\end{equation}
is the Laplacian on the tensor product space (\ref{m6defi}). The
equations of motion for the scalar fields $\varphi$ and $C_{[0]}$
and for the 3-form field strength $F^{NS}_{[3]}$ and
$F^{RR}_{[3]}$ can be better analyzed introducing a complex
notation. Defining
\begin{eqnarray}
{\mathcal{H}}_\pm & = & \pm 2 \,e^{-\varphi/2}
F^{NS}_{[3]} + {\rm i} 2 \,e^{\varphi/2} \,F^{RR}_{[3]}
\label{mcH} \\
P & =& \ft 1 2 \, d\varphi -{\rm i} \ft 12 \, e^\varphi \,
F_{[1]}^{RR} \label{Psc}
\end{eqnarray}
eq.s (\ref{NSscalapr})-(\ref{3formRR}) can be written respectively
as:
\begin{eqnarray}
d(\star P)- {\rm i} e^{\varphi} dC_{[0]}\wedge \star P + \ft 1
{16}
   {\mathcal{H}}_+ \, \wedge \, \star
  {\mathcal{H}}_+=0 \label{dstarP} \\
d \star {\mathcal{H}}_+ - \frac{{\rm i}}{2} e^{\varphi}
dC_{[0]} \wedge \, \star {\mathcal{H}}_+=
   {\rm i} \, {F}_{[5]}^{RR}\,
  \wedge \, {\mathcal{H}}_+  - P \wedge \star
  {\mathcal{H}}_- \label{hodge2formeq}
\end{eqnarray}
while eq. (\ref{f5RR}) for the 5-form becomes:
\begin{equation}
d\star F^{RR}_{[5]} = {\rm i} \, \ft 1 {8} \,
{\mathcal{H}}_+ \wedge {\mathcal{H}}_-
\label{f5equazia}
\end{equation}
\par
Besides assuming the structure (\ref{ansazzo}) we also assume that
the two scalar fields, namely the dilaton $\varphi$ and the
Ramond-Ramond $0$-form $C_{[0]}$ are constant and vanishing:
\begin{equation}
  \varphi=0 \quad ; \quad C_{[0]}=0
\label{zerodilat}
\end{equation}
As we shall see, this assumption simplifies considerably the
equations of motion.
\par
The basic ansatz  characterizing our  solution and  providing  its
interpretation as a 3-brane with a flux  through a homology
2-cycle of the ALE space is given by the following:
\begin{eqnarray}
  {\mathcal{H}}_+ & = & 2 \,d \gamma_I (z,{\bar z} )\, \wedge \,
  \omega^I \nonumber\\
  {\mathcal{H}}_- & = & -2 \, d {\bar \gamma_I} (z,{\bar z} )\, \wedge \,
  \omega^I
\label{hpmposiz}
\end{eqnarray}
where $\gamma_I (z,{\bar z} )$ is a complex field depending only
on the  coordinates $z,\bar z$ of $\mathbb{R}^2$, while $\omega^I$
($I=1,\dots,k$) constitute a basis for the space of square
integrable, anti-self-dual, harmonic forms on the ALE manifold. Notice that
upon the ansatz (\ref{zerodilat}), ${\mathcal{H}}_+$ equals the usual complex
3-form $G_{[3]}$ made up by the R-R and NS-NS type IIB 3-forms. More precisely
${\mathcal{H}}_+= 2 \,G_{[3]}$ while ${\mathcal{H}}_-= - 2 \,\bar G_{[3]}$. From
eq.(\ref{hpmposiz}) one then sees that our solutions fall in the same class
of solutions discussed in \cite{GRANA2}, dubbed type B solutions, which are dual
to M-theory solutions found by Becker and Becker in \cite{Becker}. The former are in
fact characterized by the 3-form $G_{[3]}$ being a self-dual $(2,1)$ form.
\par
As it is well known \footnote{See for instance \cite{Anselmi:1994sm}
for an early summary of ALE geometry in relation with superstrings
and conformal field theories. This relation was developed in
\cite{Douglas:1996sw,Johnson:1997py} and is of primary
relevance in connection with D-branes.} a smooth ALE manifold,
arising from the resolution of a $\mathbb{C}^2/\Gamma$ singularity,
where $\Gamma \subset \mathrm{SU(2)}$ is a discrete Kleinian
group, has Hirzebruch signature:
\begin{equation}
  \tau =\left( \# \mbox{of conjugacy classes of $\Gamma$}\right) - 1
       = \mbox{rank of } \,\mathbb{G}(\Gamma)
\label{Hirze}
\end{equation}
In the above formula $\mathbb{G}(\Gamma)$ is the simply laced Lie
algebra corresponding to $\Gamma$ in the ADE classification of
Kleinian groups, based on the Mac Kay correspondence
\cite{mackay}. As a result of eq.(\ref{Hirze}) the ALE manifold
that is HyperK\"ahler admits a triplet of self-dual $2$-forms that
are non-integrable and exactly $\tau$ integrable anti-self-dual
harmonic $2$-forms. For these latter one can choose a basis
$\omega^I$ that is dual to the integral homology basis of
$2$-cycles $\Sigma_I$ whose intersection matrix is the Cartan
matrix $\mathcal{C}$ of $\mathbb{G}(\Gamma)$.  Explicitly we can
write:
\begin{eqnarray}
\int_{\Sigma_K} \,\omega^I&=&\delta^I_K\nonumber\\
\int_{ALE} \omega^I\wedge \omega^J &=&-\,{\cal C}^{-1\, IJ}\nonumber\\
\omega^I\wedge \omega^J &=& -\,\Delta^{IJ}(y)\,\Omega_{ALE}
\label{omega}
\end{eqnarray}
where ${\cal C}^{IJ}$ is the Cartan matrix of the corresponding
(non-extended) ADE Dynkin diagram and $\Delta^{IJ}(y)$ is a
positive definite matrix whose entries are functions of the ALE
space coordinates $y$'s. The anti-self-duality of the $\omega^I$
guarantees that $\int \omega^I\wedge\star \omega^J$ is positive
definite. Finally, following the conventions of appendix
\ref{notazie} (see in particular eq.(\ref{hodge}) ) $\Omega_{ALE}$
is the volume $4$-form of the ALE manifold.

If we insert the ans\"atze (\ref{zerodilat},\ref{hpmposiz}) into
the scalar field equations (\ref{NSscalapr}, \ref{RRscalapr}) we
obtain:
\begin{eqnarray}
{\mathcal{H}}_+ \, \wedge \, \star
{\mathcal{H}}_+=0 \label{holovincol}
\end{eqnarray}
which  in turn implies that
$\partial_z\gamma_I\,\bar{\partial}_{\bar{z}}\gamma_J=0$. This
equation is solved by choosing $\gamma_I$ to be holomorphic:
\begin{equation}
  \bar{\partial} \gamma_I=0 \quad \mbox{where} \quad \partial = dz \, \frac{\partial}{\partial
  z} \quad ; \quad {\bar \partial} = d{\bar z} \, \frac{\partial}{\partial
  {\bar z}}
\label{holomgamma}
\end{equation}
Next we consider the self-dual $5$-form $F_{[5]}^{RR}$ which,
because of its definition, must satisfy the following Bianchi
identity:
\begin{equation}
d \, F_{[5]}^{RR} = {\rm i} \, \ft 1 8 \,{ \mathcal{H}}_+
\, \wedge \, { \mathcal{H}}_- \label{f5bianchi}
\end{equation}
Our ansatz for $F_{[5]}^{RR}$ is the following: (The volume forms
$\Omega$ are described in appendix A)
\begin{eqnarray}
F_{[5]}^{RR} & = & \alpha \left( U + \star U \right)  \nonumber\\
U & =  & d \left( H^{-1} \, \Omega_{\mathbb{R}^{1,3}} \right)
\label{f5ansaz}
\end{eqnarray}
where $\alpha$ is a constant to be determined later. By
construction $F_{[5]}^{RR}$ is self-dual and its equation of
motion is trivially satisfied. What is not guaranteed is that also
the Bianchi identity (\ref{f5bianchi}) is fulfilled. Imposing it,
results into a differential equation for the function $H(z,\bar z,
y)$. Using the Hodge duality relations (\ref{hodge}) listed in
appendix \ref{notazie} we easily get:
\begin{equation}
  \star U = {\rm i}\,\partial H \wedge \Omega_{ALE} -{\rm i}\,
  \bar{\partial} H \wedge
  \Omega_{ALE}+ \frac{{\rm i}}{2}\left(\hat{\star}\,d_{ALE}H\right)\wedge
  dz\wedge d\bar{z}
\label{starU}
\end{equation}
Taking the exterior derivative of (\ref{starU}) we get:
\begin{eqnarray}
 d\star U &=& 2{\rm i}\, \bar{\partial}\partial H\wedge
 \Omega_{ALE}+\frac{{\rm i}}{2}\left(d_{(ALE)}\hat{\star}\,d_{(ALE)} H\right)\wedge
  dz\wedge d\bar{z}=\nonumber\\
  &&\left(\square_{\mathbb{R}^2}\, H+\square_{ALE}\, H\right)\,\frac{{\rm i}}{2}\,d\bar{z}\wedge
  dz\wedge\Omega_{ALE}
 \label{calculo}
\end{eqnarray}
Since $d F_{[5]}^{RR} = \alpha d\star U$ and
\begin{equation}
{ \mathcal{H}}_+  \, \wedge \, {
  \mathcal{H}}_- = -  \,4\, \partial_z \gamma_I \,
  \partial_{\bar z}
  {\bar \gamma}_J \, \Delta^{IJ} \,d\bar{z}\wedge
  dz\wedge \Omega_{ALE}
\label{tempra}
\end{equation}
we conclude that the Bianchi identity (\ref{f5bianchi}) is
satisfied by our ansatz if:
\begin{equation}
  \left( \square_{\mathbb{R}^2} + \square_{ALE} \right)  H = - \frac
  {1}{\alpha} \partial_z \gamma_I \,
  \partial_{\bar z}
  {\bar \gamma}_J \, \Delta^{IJ}(y)
\label{maindiffe}
\end{equation}
This is the main differential equation the entire construction of
our sought for 3-brane solution can be reduced to. We
are going to show that the parameter $\alpha$ is determined by Einstein's
equation and fixed to $\alpha=1$.

Let us consider next the field equation for the complex
three-form, namely eq. (\ref{3formNS}) and  (\ref{3formRR}). Since
the two scalar fields are constant we have:
\begin{equation}
d \star {\mathcal{H}}_+ = {\rm i} \, F^{RR}_{[5]}\, \wedge
\, {\mathcal{H}}_+ \label{simpeq}
\end{equation}
Using our ansatz we immediately obtain:
\begin{eqnarray}
d \star {\mathcal{H}}_+ & = & -2\, {\rm i} H^{-2} dH \,
\wedge \, \partial \gamma_I \wedge \omega^I \, \wedge \,
\Omega_{\mathbb{R}^{1,3}}\,  +2 {\rm i} \, H^{-1} {\bar \partial}
\wedge \partial \gamma_I \wedge \, \Omega_{\mathbb{R}^{1,3}}
 \nonumber\\
{\rm i} \, F^{RR}_{[5]}\, \wedge \, {\mathcal{H}}_+ & = &
 -2\, \alpha {\rm i} H^{-2} dH \, \wedge \,
\partial \gamma_I \wedge
\omega^I \, \wedge \, \Omega_{\mathbb{R}^{1,3}}\, \label{pagnacco}
\end{eqnarray}
Hence if $\alpha=1$, as we have announced, the field equation for
the fields $\gamma_I(z)$ reduces to:
\begin{equation}
   {\bar \partial} \wedge \partial \gamma_I = 0 \, \Rightarrow \,
   \square_{\mathbb{R}^2} \gamma_I = 0
\label{sanssource}
\end{equation}
This equation has  to be appropriately interpreted. It says that
$\gamma_I$ are harmonic functions in two-dimensions as the real
and imaginary parts of any holomorphic function $\gamma_I (z)$
certainly are. The bulk equations do not impose any additional
constraint besides this condition of holomorphicity. However, in
presence of a boundary action for the $3$ brane, the equation
will be modified into:
\begin{equation}
  \square_{\mathbb{R}^2} \gamma_I = j_I(z)
\label{consorgo}
\end{equation}
$j_I(z)$ being a source term, typically a delta function. In this case $\gamma_I$
is fixed as:
\begin{equation}
  \gamma_I (z) = \int \, G_{\mathbb{R}^2} (z,z^\prime)\,  j_I(z^\prime) \, dz^\prime
\label{2gruna}
\end{equation}
where $G_{\mathbb{R}^2} (z,z^\prime)$ is the Green function of the
$\mathbb{R}^2$ Laplacian in complex coordinates and turns out to
be proportional to $\log z$. From this point of view one could
already expect the enhan\c{c}on mechanism to be at work in a very
similar way as in the orbifold case \cite{Bertolini:2000dk},\cite{marco}.
We will come back to this issue in section 4.
\par
In order to show that also the Einstein's equation is satisfied by
our ansatz we have to calculate the (trace subtracted) stress
energy tensor of the five and three index field strengths.
For this purpose we need the components of
$F_{[5]}^{RR}$. These are easily dealt with. Relying  on the
ansatz (\ref{f5ansaz}) and on eq.(\ref{splittoviel}) for the
vielbein we immediately get:
\begin{equation}
  F_{A_1 \dots A_5} =\left\{ \begin{array}{ccc}
    F_{i abcd} & = & \frac {1} {4!} \, f_i \, \epsilon_{abcd} \\
    F_{ j_1\dots i_5}  & = & \frac{1}{5!} \epsilon_{i j_1\dots j_5} \, f^i \\
    \mbox{otherwise} & = & 0 \
  \end{array} \right.
\label{F5intrinsic}
\end{equation}
where:
\begin{equation}
  f_i = - \alpha\, H^{-5/4} \, \partial_i H
\label{fidefi}
\end{equation}
Then by straightforward algebra we obtain:
\begin{eqnarray}
  T^{a}_{b}\left[ F_{[5]}^{RR}\right]  & \equiv & -75 \, F^{a \,\cdot\, \cdot\, \cdot \,\cdot} \,
  F_{b\,\cdot\, \cdot\, \cdot \,\cdot} = -  \frac 1 8 \, \delta^{a}_{b}
  \, f_p \, f^p \nonumber\\
  & = & - \alpha^2 \, \frac 1 8 \, \delta^{a}_{b}
  \, H^{-5/2} \partial_p H\, \partial^p H \nonumber\\
  T^{i}_{j}\left[  F_{[5]}^{RR}\right]  & \equiv & -75 \, F^{i \,\cdot\, \cdot\, \cdot \,\cdot} \,
  F_{j\,\cdot\, \cdot\, \cdot \,\cdot} = \frac 1 4\,  f^i \, f_j \, - \, \frac 1 8 \,
  \delta^i_j \, f_p \, f^p \nonumber\\
  & = & \alpha^2 \, \left[ \frac 1 4\, H^{-5/4}  \partial^i H \, \partial_j H  \, - \, \frac 1 8
  \, \delta^i_j \,  H^{-5/4}  \partial^p H \, \partial_p H \right]
\label{TofF5}
\end{eqnarray}
Let us next consider the complex $3$-forms $
{\mathcal{H}}_\pm$. With our ansatz (\ref{hpmposiz}) they
satisfy the 6-dimensional self-duality relation on $M_6$:
\begin{equation}
  \widehat{\star}_6 \, {\mathcal{H}}_\pm = \pm {\rm i} \, {\mathcal{H}}_\pm
\label{6starduals}
\end{equation}
which, in components, is rewritten as follows
\begin{equation}
  \pm {\rm i} \, {\mathcal{H}}_{\pm\vert ijk} = \frac {1}{3!}
  \, \epsilon_{ijk \ell mn } \, {\mathcal{H}}_{\pm}^{\ell mn}
\label{dualincompo}
\end{equation}
Eq.(\ref{dualincompo}) implies that:
\begin{eqnarray}
T^i_j\left[ {\mathcal{H}}\right]  & = & - \frac 9 {16}
\left( {\mathcal{H}}^{\phantom{+}i\, \cdot\, \cdot}_+ \,
{\mathcal{H}}_{- \vert j \,\cdot \, \cdot} \, + \,
{\mathcal{H}}^{\phantom{-} i \, \cdot \, \cdot}_- \,
{\mathcal{H}}_{+\vert j \,\cdot \, \cdot} \, - \frac 1 3
\, \delta^i_j \, {\mathcal{H}}^{\,\cdot\,\cdot\,\cdot}_-
\, {\mathcal{H}}_{+\vert\,\cdot\,\cdot\,\cdot} \right)
\nonumber\\
\null & = & - \frac 3 {32} \, \delta^i_j \,
{\mathcal{H}}^{\,\cdot\,\cdot\,\cdot}_- \,
{\mathcal{H}}_{+\vert\,\cdot\,\cdot\,\cdot}
\nonumber\\
& = & \frac 1 {32} H^{-3/2} \, \partial \, \gamma_I \,
\overline{\partial} \, \overline{\gamma}_J \, \Delta^{IJ}\, \delta^i_j
\label{TofHcappello}
\end{eqnarray}
where we have used the result:
\begin{eqnarray}
  {\mathcal{H}}_+ \, \wedge \, \star  {\mathcal{H}}_- & =
  &-8 \, H^{-1} \, \partial \, \gamma_I \,
\overline{\partial} \, \overline{\gamma}_J \, \Delta^{IJ} \,
\Omega_{ALE} \, \wedge \, \Omega_{ \mathbb{R}^{1,3} } \, \wedge \,
\Omega_{ \mathbb{R}^2} \nonumber\\
& \equiv & 3! \, {\mathcal{H}}^{\,\cdot\,\cdot\,\cdot}_+
\, {\mathcal{H}}_{-\vert\,\cdot\,\cdot\,\cdot} \, \times
\, \mathrm{Vol} \label{H+H-dual}
\end{eqnarray}
having named the complete volume form of the ten-dimensional space-time:
\begin{eqnarray}
  \mathrm{Vol} &=& H^{1/2} \,  \Omega_{ \mathbb{R}^{1,3} } \, \wedge
  \,\Omega_{ALE} \, \wedge \,
\Omega_{ \mathbb{R}^2} = \sqrt {-\mbox{det} g} \, dX^0 \wedge \,
\dots \, \wedge dX^9 \label{volume}
\end{eqnarray}
\par
Inserting eq.s (\ref{TofF5}), (\ref{TofHcappello}) and
(\ref{riccius}) into Einstein's eq.(\ref{einsteinequa}):
\begin{eqnarray}
  R^a_b & = &  T^{a}_{b}\left[F_{[5]}^{RR} \right]
- \frac 3 {16} \, \delta^a_b \,
  {\mathcal{H}}^{\,\cdot\,\cdot\,\cdot}_+ \,
{\mathcal{H}}_{-\vert\,\cdot\,\cdot\,\cdot} \nonumber\\
  R^i_j & = &T^{i}_{j}\left[F_{[5]}^{RR}\right] \, + \, T^{i}_{j}\left[ {\mathcal{H}}\right]
\label{neweinst}
\end{eqnarray}
we see that they are satisfied, provided
\begin{equation}
  \alpha= 1
\label{fixed}
\end{equation}
and the master equation (\ref{maindiffe}) is satisfied. This
concludes our proof that an exact $3$-brane solution with a non-trivial
flux through a homology $2$-cycle of a generic smooth ALE manifold, does indeed
exist. In the next subsection we prove that this bulk solution admits two Killing
spinors and is therefore a BPS state with $\mathcal{N}=2$ preserved
supersymmetry.

We would like to end this section by stressing a crucial
difference between our solution and the ``conifold'' case (see
\cite{Herzog:2001rz} for related discussions). In the former, the
$S^2$ cycles exist only near the apex and thus we do not have a
3-form flux at infinity, whereas in the latter, the base of the
cone supports such cycles and, as a result, the flux at infinity
is non-vanishing\footnote{We thank I.
 Klebanov for clarifying this point.}.

\subsection{Proof of bulk supersymmetry}
\label{susyprova}
As usual, in order to investigate the
supersymmetry properties of the bosonic solution we have found it
suffices to consider the supersymmetry transformation of the
fermionic fields (the gravitino and the dilatino) and impose that, for a
Killing spinor, they vanish identically on the chosen background.
By using the formulation of \cite{igorleo}, one easily gets:
\begin{eqnarray}
\delta\psi_M & = & \mathcal{D}_M\chi + \ft 1 {16} \, {\rm i}
\Gamma^{A_1 \dots A_5}  \,
F_{A_1 \dots A_5}\, V^{B}_M \, \Gamma_{B} \, \chi \,\nonumber\\
&& +\ft 1 {32} \left( - \Gamma_{BA_1\dots A_3} \, V^{B}_M+ 9
\Gamma^{A_1 A_2} V^{A_3}_M \right) \,
{\mathcal{H}}_{+\vert A_1 A_2 A_3} \, \chi^\star
\nonumber\\
\delta \, \lambda & = & -{\rm i} \, \ft 1 8 \, \Gamma^{A_1 A_2
A_3} \,{\mathcal{H}}_{+\vert A_1 A_2 A_3}\, \chi
\label{susyvaria}
\end{eqnarray}
where the supersymmetry parameter $\chi$ is a complex
ten-dimensional Weyl spinor:
\begin{equation}
  \Gamma_{11} \, \chi = \chi
\label{gamma11}
\end{equation}
and where we have already used the information that on our background
the dilaton and the Ramond scalar vanish. Using the gamma matrix
basis described in appendix \ref{notazie} (see in particular eq.s
(\ref{grossgamma},\ref{taumatte},\ref{gammezinque})) and writing
the $32$-component spinor $\chi$ as a tensor product:
\begin{equation}
  \chi = \varepsilon \,  \otimes \, \eta
\label{tensprodspin}
\end{equation}
of a $4$-component spinor $\varepsilon$,
related to the $3$-brane world volume
with an $8$-component spinor $\eta$ related to the transverse
manifold $M_6$, the dilatino transformation rule takes the following form:
\begin{equation}
  \delta \, \lambda = - {\rm i} \, \ft 1 8 \, \gamma_5 \, \varepsilon \, \otimes \,
  \tau^{ijk} \, {\mathcal{H}}_{+\vert ijk} \,
  \eta
\label{dilatinco1}
\end{equation}
Hence the first condition a Killing spinor must satisfy comes from
the vanishing of the dilatino rule and reads as follows:
\begin{equation}
 \Pi({\mathcal{H}}) \, \eta \,  \equiv \, \tau^{ijk} \, {\mathcal{H}}_{+\vert ijk} \,
  \eta= 0
\label{PiH}
\end{equation}
Using the self-duality relation (\ref{dualincompo}) we can
immediately verify that:
\begin{equation}
  \Pi({\mathcal{H}}) = \Pi({\mathcal{H}}) \, \ft 1 2 \left(
  1 + {\rm i} \, \tau_\bullet \right)
\label{chirotau}
\end{equation}
so that in order to solve eq. (\ref{PiH}) it suffices to impose
the $M_6$ chirality choice:
\begin{equation}
  {\rm i} \, \tau_\bullet \, \eta = - \eta
\label{chiro6scelta}
\end{equation}
Since $\Gamma_{11} = -{\rm i} \, \gamma_5 \, \otimes \,
\tau_\bullet$, eq. (\ref{chiro6scelta}) implies:
\begin{equation}
  \gamma_5 \, \varepsilon = \varepsilon
\label{chiro4obbligo}
\end{equation}
Inserting conditions (\ref{chiro6scelta}) and
(\ref{chiro4obbligo}) into the gravitino transformation rule,
by use of the specific structure  of $F_{A_1 \dots A_5}$ defined by
eq.(\ref{F5intrinsic}), we obtain two new conditions that must be
imposed on the Killing spinor:
\begin{eqnarray}
0 & = & \mathcal{D}_a \chi \, - \, \ft 1 8 f_i \, \gamma_a \otimes \tau^i
 \, \chi   - \ft 1 {32}
 \gamma_5 \, \gamma_a \, \Pi({\mathcal{H}})\, \chi^\star \label{kilspicond1}\\
0 & = & \mathcal{D}_j \chi \,  - \, \ft 1 8 f_i \, \tau^i
\tau_j \, \chi +  \ft 1 {32}  \tau_j \, \Pi({\mathcal{H}})\,
\chi^\star \label{kilspicon2}
\end{eqnarray}
In the above equations $f_i$ is the intrinsic derivative of the
harmonic function $H$  defined by eq. (\ref{fidefi}). Recalling
the form of the spin connection on our background (see
eq.(\ref{spinconnect})) we easily verify that eq.
(\ref{kilspicond1}) reduces to:
\begin{equation}
  \partial_a \, \varepsilon \, \otimes \, \eta \, - \, \ft 1 {32} \gamma_5 \,
  \gamma_a \, \varepsilon^\star \, \otimes \,
  \Pi({\mathcal{H}})\, \eta^\star =0
\label{redkispi1}
\end{equation}
This equation is satisfied if we choose constant supersymmetry
parameters on the $4$-dimensional world volume:
\begin{equation}
  \partial_a \, \varepsilon =0
\label{constepsi}
\end{equation}
and imposing the further condition
\begin{equation}
   \Pi({\mathcal{H}})\, \eta^\star =0
\label{PiHcondo}
\end{equation}
on the {\sl transverse} commuting spinor $\eta^\star$.
\par
Taking condition (\ref{PiHcondo}) into account and recalling once
again the form of the spin connection (\ref{spinconnect}), eq.
(\ref{kilspicon2}) becomes:
\begin{equation}
  0 = \varepsilon \, \otimes \, \left( \widehat{\mathcal{D}} + \ft 1
  8 \, H^{-1} \widehat{d} H \right) \, \eta
\label{withwarp}
\end{equation}
where the hatted operators $\widehat{\mathcal{D}}$ and
$\widehat{d}$ are the covariant and ordinary exterior derivatives
on $M_6 = \mathbb{R}^2 \times \mathrm{ALE}$ considered by itself
without the warp factor $H(z, \bar z, y)$.
\par
Introducing a rescaling factor $H^{-1/8}(z, \bar z, y)$ and a
further tensor product splitting of the $8$ component spinor
\begin{equation}
  \eta = H^{-1/8}(z, \bar z, y) \, \theta \, \otimes \, \xi
\label{splittus}
\end{equation}
into a $2$-component spinor $\theta$ relative to the manifold $
\mathbb{R}^2$ times a $4$-component spinor $\xi$ living on the ALE
manifold, eq. (\ref{withwarp}) reduces to the condition of
covariant constancy on ALE:
\begin{equation}
  0 = \widehat{\mathcal{D}} \xi
\label{withoutwarp}
\end{equation}
Collecting all the conditions we have found and using the gamma
matrix basis described in eq.s (\ref{tausplitte}) of the appendix
we can conclude that our background admits as many $4$-dimensional
supersymmetries on the world-volume of the $3$ brane as there are
simultaneous solutions of the following three equations:
\begin{eqnarray}
\theta \, \otimes \, \xi \, & = & \sigma_3 \, \theta \, \otimes \,
\overline{\gamma}_5 \, \xi
  \label{chiralconspi}\\
0 & = & \partial \gamma_I (z) \left( \sigma^1 - {\rm i} \,
\sigma^2 \right) \, \theta^\star \,
\otimes \, \omega^I_{uv} \, \overline{\gamma}^{uv}\,  \xi^\star   \label{aselfdual}\\
0 & = & \widehat{\mathcal{D}} \, \xi \label{covconstacondo}
\end{eqnarray}
where $\omega^I_{uv}$ are the intrinsic components of the
anti-self-dual normalizable $2$-forms on ALE.
\par
So far we have not used the specific geometric properties of the
ALE manifold. It is at this stage they come into the game,
playing an essential role. Squaring eq.(\ref{covconstacondo}) we
obtain the integrability condition
\begin{equation}
  - \ft 1 4 \, R^{uv} \, \overline{\gamma}_{uv} \, \xi =0
\label{integcond}
\end{equation}
where $R^{uv} \, \overline{\gamma}_{uv}$ is the curvature $2$-form
of the ALE manifold. This latter is HyperK\"ahler and as such it
has a triplet of {\sl covariantly constant self-dual} $2$-forms
$\Omega^x$, $(x=1,2,3)$ whose intrinsic components satisfy the
quaternionic algebra. This implies that the holonomy of the
manifold is $\mathrm{SU(2)}_L$ rather than $\mathrm{SO(4)} =
\mathrm{SU(2)_L }\times \mathrm{SU(2)_R}$ and that the curvature
$2$-form is {\sl anti-self-dual}. This follows from the
integrability condition for the covariant constancy of the
self-dual HyperK\"ahler $2$-forms. On the other hand, from the
Hirzebruch signature $\tau$ of the ALE manifold it follows that
there are exactly $\tau$ normalizable anti-self-dual forms
$\omega_I$. From the trivial gamma matrix identity:
\begin{equation}
  \overline{\gamma}_5 \, \overline{\gamma}_{uv} = - \ft 1 2 \,
  \epsilon_{uv st} \, \overline{\gamma}^{st}
\label{gamma5duali}
\end{equation}
follows that the two chirality eigenspaces:
\begin{equation}
  \overline{\gamma}_5 \xi_\pm = \pm \xi_\pm
\label{chirspace}
\end{equation}
are respectively annihilated by the contraction of
$\overline{\gamma}_{uv}$ with any self-dual or anti self-dual
$2$-form. Therefore it suffices to choose:
\begin{equation}
  \xi= \xi_+
\label{xi=xim}
\end{equation}
and equations (\ref{aselfdual}), (\ref{integcond}) are
simultaneously satisfied. Eq. (\ref{chiralconspi}) is satisfied by
setting:
\begin{equation}
  \sigma_3 \, \theta \, =  \theta
\label{chiralr2}
\end{equation}
and once the integrability condition (\ref{integcond}) is
fulfilled eq. (\ref{covconstacondo}) can be integrated yielding
two linear independent solutions $\xi_+^{1,2}$ spanning the
irreducible representation $(0,1/2)$ of $\mathrm{SO(4)}$. The
other irreducible representation $(1/2,0)$ corresponds to spinors
that are not Killing and do not generate supersymmetries preserved
by the background. In conclusion we have $2$ Killing spinors
generating an $\mathcal{N}=2$ supersymmetry on the world volume.
In other words the bosonic background we have constructed
corresponds to a BPS state preserving a total of $2 \times 4 = 8$
supercharges.

\section{The Eguchi-Hanson case}
\label{Ehcase}
As showed in the previous sections, the complete
integration of the supergravity field equations is reduced
to the solution of a single differential equation, namely eq.
(\ref{maindiffe}). In this section we investigate
the properties of such an equation, choosing the simplest instance
of an  ALE manifold, namely the Eguchi-Hanson space
\cite{Eguchi:1978xp}, which in the ADE classification corresponds to
$A_1 \sim \mathbb{Z}_2$ (see appendix \ref{EH} for a more detailed
treatment of the Eguchi-Hanson space and for the definition of
the relevant conventions). While in this section we will go through a somewhat
detailed mathematical analysis, section 5 contains a summary of the
most relevant results and their physical implications.
It should be said from the beginning that, whereas section 3 is entirely based
on type IIB supergravity, in the following we are going to make some (we
believe resonable) assumptions on the nature of the microscopic theory,
namely the structure of the source terms needed to fix the boundary
conditions.

The Eguchi-Hanson metric has the form:
\begin{eqnarray}
ds^2_{EH}&=& g(r)^{-2}\,dr^2+ r^2\,\left(\sigma_x^2+
\sigma_y^2\right)+r^2\,g(r)^2\,
\sigma_z^2\nonumber\\
g(r)^2&=&1-\left(\frac{a}{r}\right)^4 \label{eguchi}
\end{eqnarray}
where $r\ge a$, $\theta\in \left[0,\,\pi\right[$ and $\psi$, $\phi
\in \left[0,\,2\pi\right[$. The $1$-forms $\sigma^i=\left\{
\sigma^x, \, \sigma^y ,\, \sigma^z \right \}$ are listed in eq.
(\ref{sigmai}) of appendix \ref{EH}.\par This space has a unique
homology 2-cycle $\Sigma$ located at $r=a$ and spanned by the
coordinates $\theta$, $\phi$. The anti-self dual form $\omega$
fulfilling eqs. (\ref{omega}) is:
\begin{eqnarray}
\omega &=&\frac{a^2}{2\,\pi}\,d\left(\frac{\sigma_z}{r^2}\right)
\end{eqnarray}
The function $\Delta(y)$ defined in (\ref{omega}) is explicitly
evaluated to be:
\begin{eqnarray}
\Delta(y)&=&\frac{2\, a^4}{\pi^2\,r^8}\label{Delta}
\end{eqnarray}

The equation for $H$ in the Eguchi-Hanson case can then
be easily obtained from
the general expression in eq.(\ref{maindiffe}). As it is usually the case,
we make a spherically symmetric ansatz, compatible with the background at hand:
for the case we shall be interested in, the
coefficients of the equation for $H$ depend only on the
radial coordinates $\rho=\sqrt{z\bar{z}}$ and $r$ on
$\mathbb{R}^2$ and on the Eguchi-Hanson space respectively,
and we can then assume the same property for $H$ to hold and write it as
$H(\rho,\,r)$\footnote{We shall consider the case in which
$\gamma(z)=\log{(z)}$. In a more general situation
$\vert\partial\gamma(z)\vert^2$ could also depend on the angular
coordinate on $\mathbb{R}^2$. In this case the function $H$ would
have an angular dependence in $\mathbb{R}^2$ as well.}.
The equation for $H$ can then be written as:
\begin{eqnarray}
\left(\partial_4^2+\partial_5^2\right)\,H+\frac{1}{r^3}\partial_r\left(r^3\,g(r)^2\,\partial_r
H\right)&=& -\frac{2\,
a^4}{\pi^2\,r^8}\,|\partial_z\,\gamma|^2+{\cal S}(\rho,\,r)
\label{maindifEH}
\end{eqnarray}
where ${\cal S}(\rho,\,r)$ is a source term for the $3$-brane
charge for which we also make a spherical ansatz. Differently from the first
term on the right hand side, ${\cal S}(\rho,\,r)$ is not deduced from the
dynamics in the bulk. In the present analysis its presence should be intended
only for the sake of fixing boundary conditions near the cycle and it will
be determined in the following sections.

\subsection{Solution method by partial Fourier transform}
\label{partfourier}
The method we use to study eq.(\ref{maindifEH}) is based on a partial
Fourier transform on the coordinates of the flat $\mathbb{R}^2$ space
that reduces our problem to that of a massive Laplace equation on the residual
four-dimensional manifold in the unknown function $\widetilde{H}(\mu,r)$, Fourier
transform in $\mathbb{R}^2$ of $H(\rho,r) - 1$ (with $\mu$ being the modulo of the
momentum in $\mathbb{R}^2$). We refer for the details of the analysis to appendix
\ref{diffeq}. As we are going to show this method allows a parallel treatment of our
case $\mathbb{R}^2 \times \mathrm{ALE_{EH}}$ with the case $
\mathbb{R}^2 \times \mathbb{R}^4 $ which is the covering space for
fractional D3 branes on orbifolds \cite{Bertolini:2000dk}. Such a
parallelism is very instructive since the comparison with the
flat case provides the necessary insight to discuss and interpret the
appropriate boundary conditions. The main goal of the analysis in
the following sub-sections is to determine  $H(\rho,r)$ through its
near-cycle and far-from-cycle asymptotic limits and an exact
analytical description through a power series expansion of
$H$ in an open neighborhood of the cycle,
consistently with the boundary conditions.
The reason why we choose to study the equation in the Fourier transform
$\widetilde{H}(\mu,r)$ of $H(\rho,r) - 1$ instead of
$H(\rho,r)$ is a matter of convenience in dealing with the
boundary conditions at infinity: at large distances $ H(\rho,r) \, \rightarrow\,1$
and therefore $\widetilde{H}(\mu,r)\, \rightarrow\,0$.
The relation between $\widetilde{H}(\mu,r)$ and $H(\rho,r)$ is deduced from
the standard definition of Fourier transform given in appendix \ref{diffeq} and reads:
\begin{equation}
  \widetilde{H}(\mu,r) =\frac{1}{2\pi}\int_0^{\infty}
 d\rho \,\rho\, J_0(\mu\rho) \, \left(H(\rho,r)-1\right)
\label{hspecial}
\end{equation}
The equation for $\widetilde{H}(\mu,\,r)$ has the following form:
\begin{equation}
  L_r[a,\mu] \, \widetilde{H}(\mu,\,r) = J(\mu,\,r)
\label{formalform}
\end{equation}
where we have introduced the second order differential operator
\begin{eqnarray}
L_r[a,\mu] \, u &=&
  \frac{1}{r^3} \frac{d}{dr} \, \left( r^3 \, g(r)^2 \, \frac{du}{dr}
  \right)  - \mu^2 \, u
\label{diffoperat}
\end{eqnarray}
and where we have defined the source function:
\begin{equation}
  J(\mu,\,r)  \, \equiv \, -\frac{2 a^4}{\pi^2 \,
  r^8}\, j(\mu)+ S(\mu,\,r)
\label{source}
\end{equation}
$j(\mu)$ denoting the Fourier transform of
$\vert\partial\gamma(z)\vert^2$ and $S(\mu,\,r)$ being the
transform of the source term ${\cal S}(\rho,\,r)$ for the
$3$-brane charge.

The notation $L_r[a,\mu]$ is very convenient since it encodes various cases
into a single formula. For $a=0$, $L_r[0,\mu]$ is just the {\sl
massive} radial Laplacian on flat $\mathbb{R}^4$ space.
Alternatively for $\mu = 0$, $L_r[a,0]$ is the {\sl massless}
radial Laplacian on either $\mathrm{ALE_{EH}}$ or $\mathbb{R}^4$
depending on the value of $a$. By massive or massless Laplacian we
mean the operator $(\square -\mu^2)$ with $\mu^2 >0$ or
$\mu^2=0$.\par It is known from the standard theory of differential
equations that the general integral $\widetilde{H}(\mu,\,r)$ of
eq. (\ref{formalform}) has the form:
\begin{equation}
\widetilde{H}(\mu,\,r) = \beta_1\, u_1(\mu,\,r) + \beta_2\, u_2(\mu,\,r) + u_{n-h}(\mu,\,r)
\label{genintegra}
\end{equation}
where $u_{1,2}(\mu,\,r)$ are two independent solutions of the
homogeneous equation associated with (\ref{formalform}):
\begin{equation}
L_r[a,\mu] \, u =0 \label{hom}
\end{equation}
while $u_{n-h}(\mu,\,r)$ denotes a particular solution of the
non-homogeneous equation. The constants $\beta_1 $ and $\beta_2$
are fixed by the boundary conditions.

It is useful to express the unique solution of eq.(\ref{formalform}) fulfilling the
required boundary conditions in terms of a Green function
$G(a,\mu|r,\,r^\prime)$:
\begin{equation}
  \widetilde{H}(\mu,\,r) = \int_a^\infty \, G( a,\mu\vert r,r^\prime)
  J \left (\mu,\,r^\prime \right ) \,\left( r^\prime\right) ^3\, dr^\prime
\label{gensolgrun}
\end{equation}
where the Green function is defined by the equation:
\begin{equation}
  L_r[a,\mu] \,  G\left( a,\mu\vert r,r^\prime \right) = \frac{\delta(r-r^\prime)}{(r^\prime)^3}
\label{grunadefi}
\end{equation}
The defining properties of $G(a,\mu|r,\,r^\prime)$ are recalled
and thoroughly discussed in appendix \ref{diffeq}. The Green
function associated with the homogeneous equation (\ref{hom}) is
generally expressed in terms of a combination of the solutions
$u_{1,2}$. The main problem to be dealt with in the next sections
is to determine, on one hand, the source term $S(\mu,\,r)$, and, on
the other hand, the form of the Green function such that it
implements the required physical boundary conditions of the
problem. These are conditions on the behavior of $\widetilde{H}$
near the cycle $r\rightarrow a$ and at infinity, $r\rightarrow
\infty$.

\subsection{Asymptotics of the the two independent solutions
$u_1,u_2$ for the homogeneous equation}
\label{asymptsect} So far our discussion of the differential
equation has been fully general and applies to both $\mathbb{R}^4$
and to the Eguchi-Hanson case. We turn next to a more detailed
consideration of the homogeneous equation and of its basis of
solutions in various sub-cases. Indeed the parameter space of our
differential operator, spanned by $a$ and $\mu$ is partitioned
into four sub-cases according to the following scheme:
\begin{center}
  \begin{tabular}{|c||c|c|}
    \hline
    \null & $a=0$ & $a \ne 0 $ \\
    \hline
    \hline
    $\mu=0$ & massless $\mathbb{R}^4$ case& massless $EH$  case\\
    \hline
    $\mu \ne 0$ & massive $\mathbb{R}^4$ & massive $EH$  case \\ \hline
  \end{tabular}
\end{center}
In three of the above four sectors the homogeneous equation can be
solved in terms of elementary transcendental or special functions
so that all asymptotic behaviors are easily calculated. The really
hard corner is the lowest right one, namely the massive
Eguchi-Hanson case. As we show in the next section, for
$\mu\ne0,a\ne0$ the homogeneous equation (\ref{formalform}) is a
particular instance of the confluent Heun differential equation
and its solutions can be expressed in terms of a
power series. The same pattern of asymptotic behavior is
reproduced also in this case and provides the needed guideline to
fix the boundary conditions.

Let us write down the basis of solutions in the three cases where
they are easily accessible by elementary methods and indicate our
nomenclature for the massive EH case that we address in the
following section:
\begin{equation}
  \begin{array}{ccccc}
      a=0 \, ; \,\mu=0 &\mbox{ \underline{massless $\mathbb{R}^4$}} & u_1(0,0,r) & = & \frac{1}{r^2} \\
      \null & \null & u_2(0,0,r) & = & 1 \\
      a=0 \, ; \, \mu \ne0 ;& \mbox{ \underline{massive $\mathbb{R}^4$}} & u_1(0,\mu,r) & = &
      \frac{\mu}{r}\,K_1\left(\mu \, r\right) \\
      \null & \null &  u_2(0,\mu,r) & = & \frac{2}{\mu r} \, I_1(\mu r) \\
      a\ne0 \, ; \,\mu=0 & \mbox{ \underline{massless $EH$}} & u_1(a,0,r) & = & -\frac{1}{2\,a^2}\,
      {\rm log} \left( \frac{r^2-a^2}{r^2+a^2}\right)  \\
      \null & \null & u_2(a,0,r) & = & 1 \\
      a\ne0 \, ; \,\mu\ne0 & \mbox{ \underline{massive $EH$}} & u_1(a,\mu,r) & = &-\frac{1}{2 \, a^2}\, H^{(I)}(\mu,r) \\
      \null & \null & u_2(a,\mu,r) & = & H^{(R)}(\mu,r) \\
    \end{array}
\label{sforzo}
\end{equation}
where $K_\nu(z)$ and $I_\nu(z)$ denote Bessel function of index
$\nu$ and type $K$ or $I$, respectively. The reader can easily
check that the above solutions have been normalized in such a way
as to fulfill the normalization  (\ref{fissonorma}) of the Wronskian.
The notation $H^{(I)}$ and
$H^{(R)}$ is reserved for the confluent Heun functions, to be
discussed in the next section, that have the corresponding
behavior at $r \to a$ as their analogues in the massless case. To
understand the rationale of the previous statement it is
instructive to consider the series expansion of $u_{1,2}(0,\mu,r)$
at $r=0$. We find:
\begin{eqnarray}
u_1(0,\mu,r) & \simeq_{\,\, r\to 0} & \frac{1}{r^{2}} +
\frac{\mu^2}{4}\, \left(-1+2\gamma +2
\log{(\frac{r\,\mu}{2})}\right) + \mathcal{O}(r^2)
\label{u1asympto}\\
  \null & \null & \null \nonumber\\
u_2(0,\mu,r) & \simeq_{\,\, r\to 0} &1 + \frac{{\mu }^2\,r^2}{8}
+\mathcal{O}(r^4)
\label{u2asympto}
\end{eqnarray}
where $\gamma=0.577216$ is the Euler gamma number. As one sees the
leading behavior of the massive solutions for $r \to 0$ is just
identical with the behavior of the massless solutions, $u_1$ is
singular and has a second order pole $1/r^2$ while $u_2$ is a
regular power series and tends to $1$. Indeed choosing the
overall normalization consistent with eq. (\ref{fissonorma}) we
could also arrange for the coefficients of the pole and of the
constant term in $u_1$ and $u_2$ respectively, to be equal to one.
In the case of the regular solution $u_2(0,\mu,r)$ the generic
coefficient of the power series expansion in eq.(\ref{u2asympto})
can be easily calculated and we have:
\begin{equation}
  u_2(0,\mu,r) = \sum_{n=0}^{\infty} \,
  \frac{1 + n}{4^n\, ((1 + n)!)^2} \, \left( \mu r\right)^{2n}
\label{u2asymptform}
\end{equation}
As we see in the next section a similar regular series, although
much more complicated can be written down for $ u_2(a,\mu,r)\equiv
h^{reg}(a,\mu,r)$. Let us next consider the behavior at $r=a$ of
the massless solution $u_2(a,0,r)$. We have
\begin{eqnarray}
u_1(a,0,r) & \simeq_{\,\, r\to a} &
\frac{-1}{2\,a^2}\,\log{\left(\frac{r-a}{a}\right)}+\mathcal{O}(r-a)
\label{nomassEHasym}
\end{eqnarray}
As we see the double pole  singularity $1/r^2$ appearing in the
$\mathbb{R}^4$ case has been replaced by a logarithmic singularity
$\log (r-a)$ appearing in the Eguchi-Hanson case. What remains
identical in the two cases is the relation between the massive and
massless solutions. Also for $\mu \ne 0$ we find in the next
section that:
\begin{eqnarray}
u_1(a,\mu,r) & \simeq_{\,\, r\to a} &  \frac{-
       \log ( r-a)  }{2\,a^2} + \mbox{reg.} \nonumber\\
u_2(a,\mu,r) &\simeq_{\,\, r\to a} & 1 + \mathcal{O}(r-a)
\label{andamenta}
\end{eqnarray}
Using this information on the asymptotic behavior of the two
fundamental solutions we can complete our program of
constructing the Green function by imposing the boundary
conditions of physical relevance.

To this effect we need the preliminary analysis of the
near-cycle limit in the Eguchi-Hanson space. Let us go back to the
Eguchi-Hanson metric in the form (\ref{eguchi}) and let us perform
the following change of variable:
\begin{equation}
 r\rightarrow v  =
    \frac{{\sqrt{ r^4-a^4 }}}{2\,r}
\label{cambioadu}
\end{equation}
 The form of the metric near the cycle $(r \to a)$ can
now be calculated by expanding the metric (\ref{eguchi}) in
power series of $v$ at $v=0$. We obtain:
\begin{eqnarray}
ds^2_{EH} & \simeq_{v \to 0} & dv^2 +v^2\,d\psi^2
 + \frac{a^2}{4} \, \left ( d\theta^2
   \, +  \,
     d\phi^2\,\sin^2 \theta  \right ) +\mathcal{O}\left(v^4\right)
\label{ubipicchio}
\end{eqnarray}
showing that near the homology cycle the Eguchi-Hanson metric
approximates that of a manifold $ \mathbb{R}^2 \, \times \,
\mathrm{S}^2$.

In view of this result we consider the transcription of the
homogeneous radial equation to the new variable $v$, defined
by eq.(\ref{cambioadu}). We obtain, for small $v$:
\begin{equation}
   \frac{d^2}{dv^2}  \,
  \widetilde{H} + \frac{1}{v} \, \frac{d}{dv} \, \widetilde{H} - \mu^2 \,
  \widetilde{H} = 0
\label{prucio}
\end{equation}
admitting the following two linearly independent solutions:
\begin{equation}
  \overline{u}_1(v,\mu) = \frac{1}{a^2} \, K_0(v\, \mu) \quad ; \quad
  \overline{u}_2(v,\mu) =  \, I_0(v\, \mu)
\label{ubarsolv}
\end{equation}
whose behavior for $v \to 0$ is:
\begin{eqnarray}
 \overline{u}_1(v,\mu) & \simeq_{v\to 0} & -\frac{\log(v)}{a^2} +\mbox{reg}\nonumber\\
 \overline{u}_2(v,\mu) & \simeq_{v\to 0} &1 + \frac{{\mu }^2\,v^2}{4} +
  {\mathcal{O}(v^3)}
\label{ubarrilimi}
\end{eqnarray}
If in eq.s (\ref{ubarrilimi}) we reinsert the original variable
$r$ via the relation (\ref{cambioadu}) we obtain the announced
proof that, near the cycle, the leading behavior of the massive
solutions $u_1(a,\mu,r)$  is the same as the behavior of the
corresponding massless solutions. By such a substitution we
get:
\begin{equation}
\begin{array}{rclcl}
u_1(a,\mu,r) &  \simeq_{r\to a} & \overline{u}_1(
\frac{\sqrt{(r-a)}}{a},\mu)&=& -\frac{1}{2 a^2}\log (r-a) +\mbox{reg}\\
u_2(a,\mu,r) &  \simeq_{r\to a} &
\overline{u}_2(\frac{\sqrt{(r-a)}}{a},\mu) & = & 1+
\mathcal{O}(r-a)\
\end{array}
\label{andamenti}
\end{equation}

\subsection{The $3$-brane charge, the physical boundary
conditions and the final form of the Green function}
\label{d3branacargo}
Equipped with the results of the previous
section, let us now consider the case of a $3$-brane with vanishing 3-form
placed either at the origin of $ \mathbb{R}^6$ in the orbifold
case where the transverse space is $M_6 =\mathbb{R}^2 \times
\mathbb{R}^4/\Gamma$  or at the homology cycle $r=a$ in the case
where the transverse space is $M_6 =\mathbb{R}^2 \times
\mathrm{ALE_{EH}}$. Naming $Q$ the charge of such a brane, the
expected behavior of the $H$ function near the brane is, in the
two cases, the following one:
\begin{equation}
H \simeq \frac{Q}{(x_4^2 + x_5^2 + \dots+x_9^2)^2}
\quad\mbox{for $ \mathbb{R}^6$; }\quad
H \simeq  \frac{Q/{2 a^2 }}{\, (\rho^2 + v^2)}\quad
\mbox{for $ \mathbb{R}^2 \times$ ALE }
\label{rpowers}
\end{equation}
The first of eq.s (\ref{rpowers}) is obvious. The second is due to
the discussion of the previous section. Near the cycle
the Laplacian on $M_6$ becomes:
\begin{equation}
  \square_{M_6} \sim \square_{\mathbb{R}^4} +\square_{S^2}
\label{lilloma}
\end{equation}
and since $S^2$ is a compact, positive curvature manifold there
are no zero modes of $\square_{S^2}$ except the constant.
Therefore the non trivial part of $H$ behaves as a harmonic
function on $\mathbb{R}^4$, namely the second of eq.s
(\ref{rpowers}). The  scale factor $1/(a^2)$ appearing there
is understood as follows. If $Q$ is
the total charge perceived at infinity, the density of charge on
the homology $2$-sphere of radius $\frac a 2$ is:
\begin{equation}
  q=\frac{Q}{a^2 \pi}
\label{chargedensity}
\end{equation}
Hence what appears as charge in the near cycle $\mathbb{R}^4$
plane is $q$ rather than $Q$. Finally the $\pi/2$ factor that  is
needed to match the second with the first of equations
(\ref{rpowers}) is just a matter of convenient normalization. Let
us now perform the Fourier-Bessel transform of eq.s
(\ref{rpowers}). We obtain
\begin{eqnarray}
\frac{1}{2\pi} \, \int_0^\infty  \, J_0(\mu\,\rho) \, \rho \,
\frac{Q}{(r^2+\rho^2)^2} \, d\rho & = & \frac{Q}{4 \pi} \,
\frac{\mu}{r} \, K_1\left (\mu\,  r\right) =
\frac{Q}{4 \pi} \, u_1(0,\mu,r) \nonumber\\
\frac{1}{2\pi} \, \int_0^\infty  \, J_0(\mu\,\rho) \, \rho \,
\frac{Q}{2 a^2 \, \left (v^2+\rho^2 \right )} d\rho & = &
  \frac{Q}{4 \pi} \,
\frac{1}{a^2 }  \, K_0\left (\mu \, v\right) = \frac{Q}{4 \pi}
\overline{u}_1(v,\mu) \label{Qbrancargo}
\end{eqnarray}
The important conclusion implied by the above analysis is that, in
terms of the fundamental basis of solutions $u_1,u_2$ we have
obtained the physically appropriate boundary conditions for the
function $\widetilde{H}(\mu,r)$. In both the orbifold or smooth
ALE case, in the limits $r\rightarrow 0$ and $r\rightarrow a$
respectively, we have:
\begin{eqnarray}
  \widetilde{H}(\mu,r) & \sim &
  \frac{Q}{4 \pi}  {u}^{div}_1(\mu,\,r)
  + \mbox{reg}
\label{divergent}
\end{eqnarray}
where $Q$ is the $3$-brane charge and ${u}^{div}_1(\mu,\,r)$
denotes the divergent part of the solution $u_1(a,\,\mu,\,r)$
defined in eq.s(\ref{sforzo}). This condition, together with the
boundary condition at infinity,  fixes the coefficients
$\beta_1,\,\beta_2$ in eq. (\ref{genintegra}) to be:
\begin{equation}
  \beta_1 = \frac{Q}{4 \pi} \quad ; \quad \beta_2 = 0
\label{albetvalues}
\end{equation}
The appropriate expression for the Green function in order for it
to implement the physical boundary conditions defined above is
derived in appendix \ref{diffeq} and is the following:
\begin{equation}
G(r,r^\prime) = \cases {- \ft 1 2
  u_2(r) \, u_1(r^\prime ) \quad \mbox{for} \, r<r^\prime \cr
  - \ft 1 2
  u_1(r) \, u_2(r^\prime  ) \quad \mbox{for} \, r>r^\prime\cr}
\label{trulagruna}
\end{equation}
Acting with such a Green function on a source:
\begin{equation}
S(\mu,\,r) = - \frac{Q}{2 \pi} \, \frac{\delta(r-a)}{r^3}
\label{risorgo}
\end{equation}
we obtain a solution with the prescribed behavior as in eq.
(\ref{divergent}). From the physical viewpoint the source
(\ref{risorgo}) comes from the world-volume action of the
$3$-brane.

\subsection{Reduction to the confluent Heun equation}
\label{heunsoft}
As we have seen in previous sections a convenient
approach to the solution of  equation (\ref{maindifEH}) relies on
the partial Fourier transform of eq.(\ref{genfourier}) leading to
the new eq.s (\ref{formalform}), (\ref{diffoperat}) for the function
$\widetilde{H}(\mu,r)$. So far we have analyzed such an equation
from a general viewpoint and we have discussed how to impose the
correct boundary conditions from the asymptotic behavior of the
two solutions $u_1$ and $u_2$ of the associated homogeneous
equation. In this section we eventually come to a more in depth
analysis of the  equation in a generic point of its moduli space,
i.e. at $\mu \ne 0, a\ne 0$. \par Parameterizing the radial
direction in the Eguchi-Hanson space through the new variable
$w=(a^2-r^2)/(2 a^2)$, equation
(\ref{formalform}),(\ref{diffoperat}) can be rewritten in the
following form:
\begin{eqnarray}
w^2\,\frac{d^2 \widetilde{H}}{dw^2}+w\,P_1(w)\,\frac{d
\widetilde{H}}{dw}+ P_2(\mu,\,w)\, \widetilde{H}&=&
\frac{h(\mu)\, w}{(1-w)(1-2w)^3}\nonumber\\
\widetilde{H}(\mu,w)&\stackrel{ w\rightarrow
-\infty}{\longrightarrow}& 0
 \label{heun}
\end{eqnarray}
where
\begin{equation}
\begin{array}{rclcrcl}
P_1(w)&=&\frac{2w-1}{w-1} & ; &
P_2(\mu,\,w)&=& k\,w\,\left(\frac{2w-1}{w-1}\right) \\
h(\mu)&=&\frac{1}{2\pi^2\, a^{2}}\,j(\mu)=\frac{1}{2 \pi^2
a^2}\int_0^\infty\, \frac{d^2 x}{(2\pi)^2}\,\vert\partial_z
\gamma(z)\vert^2\,e^{{\rm i}\vec{p}\cdot \vec{x}} & ; & k&=&
\frac{\mu^2 a^2}{4}\
\end{array}
\label{perfectus}
\end{equation}
We have omitted in the source term the part proportional to
$S(\mu,\,r)$ since, as explained in the previous sections, its
effect amounts in a determination of the relative coefficients of
the two independent solutions $u_{1,2}$ of the homogeneous
equation, as in eq. (\ref{albetvalues}). We postpone to appendix
a more detailed analysis of eq. (\ref{heun}), as a
non-homogeneous confluent Heun equation, and of its solution. In
the present section we shall determine the analytical expression
of $\widetilde{H}$ in a finite neighborhood of the cycle,
represented by the regular singularity $w=0$ and implement on it
the the required boundary conditions: the behavior of the integral
at infinity, defined in eq. (\ref{heun}), and its behavior at
$w\rightarrow 0$, determined by the presence of source terms as we
explained in section \ref{d3branacargo}. Finally, using these
exact results  we shall compute the anti-Fourier transform $H$ of
$\widetilde{H}$ in its asymptotic near-cycle limit, confirming
the predictions made in the previous section.

Let us now discuss the solution around $w=0$ and the
asymptotic solution for $w\rightarrow -\infty$.
The expansion of the solution around $w=0$ corresponds to
considering its behavior  near the exceptional cycle: $r\sim
a$. Since $w=0 $ is a regular singularity there are standard ways
to determine the power series expansion of the the solutions for
both the homogeneous \cite{dennery} and non-homogeneous equations
around such a point.

First let us focus on the homogeneous equation associated with
(\ref{heun}) and let us denote it symbolically by $L_w\,\widetilde{H}=0$.
This equation has two independent solutions, one being regular at
$w=0$ will be denoted by $H^{(R)}(\mu,\,w)$, the other having
a logarithmic behavior near the same point will be denoted by
$H^{(I)}(\mu,\,w)$. From standard analysis we may determine the
power series expansions of the above integrals around $w=0$:
\begin{eqnarray}
H^{(R)}(\mu,\,w)&=&\sum_{n=0}^{\infty}\, c^R_n(\mu)\,
w^n=\sum_{n=0}^{\infty}\, c_n(\mu,\,\sigma=0)\, w^n\nonumber\\
H^{(I)}(\mu,\,w)&=&\lim_{\sigma\rightarrow
0}\,\frac{d}{d\sigma}\left(w^\sigma\, \sum_{n=0}^{\infty}\,
c_n(\mu,\,\sigma)\, w^n \right) \label{hir}
\end{eqnarray}
the coefficients $c_n(\mu,\,\sigma)$ will be derived in appendix
\ref{diffeq} and have the form:
\begin{eqnarray}
c_n(\mu,\,\sigma) &=&(-)^n\,
c_0(\mu)\,\frac{F_n(\mu,\,\sigma)}{\prod_{j=1}^{n}\,
f_0(\sigma+j)}=(-)^n\,
c_0(\mu)\,\frac{F_n(\mu,\,\sigma)}{\prod_{j=1}^{n}\,
(\sigma+j)^2}\nonumber\\
&&\nonumber\\
F_n(\mu,\,\sigma)&=&{\rm det}\,\mathbb{F}_n(\mu,\,\sigma)
\label{coefreg}
\end{eqnarray}
where $\mathbb{F}_n(\mu,\,\sigma)$ is an $n\times n$ matrix, see
appendix \ref{diffeq}, whose entries are:
\begin{eqnarray}
\left[\mathbb{F}_n(\mu,\,\sigma)\right]_{i\,i}&=& k-\sigma
-n+i\,;\qquad \left[\mathbb{F}_n(\mu,\,\sigma)\right]_{i\,j>i}=
-(k+\sigma+n-j)\nonumber\\
\left[\mathbb{F}_n(\mu,\,\sigma)\right]_{i+1\,i}&=&(\sigma+n-i)^2\,;\qquad
i,\,j=1,\dots, n
\end{eqnarray}
Both the series for $H^{(I)}(\mu,\,w)$ and $H^{(R)}(\mu,\,w)$
converge for $|w|<1$, namely for $r^2-a^2< 2 a^2$. A generic
solution of the homogeneous equation will be expressed as a linear
combination of $H^{(I)}(\mu,\,w)$ and $H^{(R)}(\mu,\,w)$, the
relative coefficients will be determined in quite the same way as
the constants $\beta_1,\,\beta_2$ defined by eq.
(\ref{genintegra}) were determined in section \ref{d3branacargo}.

Let us now consider a particular series solution of the full
non-homogeneous equation
\begin{equation}
H^{(n-h)}(\mu,\,w)=\sum_{n=0}^{\infty}\, c^{n-h}_n\, w^n
\label{particula}
\end{equation}
the coefficients $c^{n-h}_n$ are derived in appendix
\ref{diffeq} and have the form
\begin{eqnarray}
c_n^{(n-h)}(\mu)
&=&(-)^n\,\frac{\tilde{F}_n(\mu)}{\prod_{j=1}^{n}\,
f_0(j)}=(-)^n\,\frac{\tilde{F}_n(\mu)}{\prod_{j=1}^{n}\,
j^2}\nonumber\\
\tilde{F}_n(\mu)&=&{\rm det}\,\tilde{\mathbb{F}}_n(\mu)
\label{coefih}
\end{eqnarray}
where $\tilde{\mathbb{F}}_n(\mu)$ is an $n\times n$ matrix similar
in form to $\mathbb{F}_n(\mu,\,\sigma)$ except for the last column
which depends on the source term expansion:
\begin{eqnarray}
\left[\mathbb{F}_n(\mu,\,\sigma)\right]_{i\,i}&=& k-\sigma
-n+i\,;\qquad \left[\mathbb{F}_n(\mu,\,\sigma)\right]_{i\,j>i}=
-(k+\sigma+n-j)\nonumber\\
\left[\mathbb{F}_n(\mu,\,\sigma)\right]_{i+1\,i}& = &(\sigma+n-i)^2\,; \nonumber\\
&& \left[\mathbb{F}_n(\mu,\,\sigma)\right]_{i\,n}=-h(\mu)\,s_{n-i}\,;\qquad
i=1,\dots, n\,;\,\,j=1,\dots,n-1
\label{sorgen}
\end{eqnarray}
 where the coefficients $s_i$ are defined by the power series expansion of the source term
  on the right hand side of eq. (\ref{heun}):
\begin{eqnarray}
\frac{h(\mu) \,w}{(1-w)(1-2w)^3}&=& h(\mu)\, \sum_{n=0}^{\infty}\,
s_{n}\,
w^{n+1}\nonumber\\
s_n&=&\sum_{i=0}^n\, (i+1)(i+2)2^{i-1} \label{ss}
\end{eqnarray}
it can be checked that both the series expansions for the source
term and for $H^{(n-h)}(\mu,\,w)$ converge for $|w|<1/2$, i.e. for
$r^2-a^2<a^2$.
\par Below we write the expansions of
$H^{(R)}(\mu,\,w)$, $H^{(I)}(\mu,\,w)$ and $H^{(n-h)}(\mu,\,w)$ up
to order $w^3$:
\begin{eqnarray}
H^{(R)}(\mu,\,w)&=&c_0^R(\mu)\,\left(1-k w+\frac{k^2}{4}\, w^2+O(w^3)\right)\nonumber\\
H^{(I)}(\mu,\,w)&=&c_0^I(\mu)\,\left(\log{(w)}+\left(-k
\log{(w)}+1+2k\right)\, w \right. \nonumber\\ && \left. +\,
\frac{1}{4}\left(k^2\log{(w)}+2-3 k^2\right)\, w^2 +
\dots\right)\nonumber\\
H^{(n-h)}(\mu,\,w)&=& h(\mu)\, \left(w+\frac{1}{4}\,(8-k)
w^2+O(w^3)\right) \label{ord3}
\end{eqnarray}
In some particular limits  the above series can be summed.\par
First of all let us consider the solution at $\rho\gg a$ to which
only the Fourier transform $\tilde{H}$ at $\mu=0$ ($k=0$)
contributes. In this case, the series  in eqs. (\ref{ord3}) are
easily summed and we get:
\begin{eqnarray}
H^{(R)}(\mu,\,w)&=& 1\label{u2lim}\\
H^{(I)}(\mu,\,w)&=&\log{(w)}+\sum_{n=1}^{\infty}\,
\frac{w^n}{n}=\log{(-w)}-\log{(1-w)}+\mbox{const}.\nonumber\\
&=&
\log{\left(\frac{r^2-a^2}{r^2+a^2}\right)}+\mbox{const}.\label{u1lim}\\
H^{(n-h)}(\mu,\,w)&=& \frac{h(\mu)}{2 (1-2 w)}-\frac{h(\mu)}{2}=
\frac{a^2\,h(\mu)}{2 r^2}+\mbox{const}. \label{k0}
\end{eqnarray}
Eq.s (\ref{u2lim}) and (\ref{u1lim}) are a very nice confirmation
of what we had already established in previous sections
considering the limiting form of the equation either for $\mu \to
0$ or for $a\to 0$. In particular eq.s (\ref{u2lim}) and
(\ref{u1lim}) justify the absolute normalization of the solutions
$u_{1,2}$ chosen in eq.(\ref{sforzo}).\par Another accessible
limit is that corresponding to $\mu^2 \gg 1/a^2$ with $|w|\ll 1$
(i.e. $r^2-a^2 \ll a^2$) and  keeping $|k\,w|<1$. This is the
asymptotic near-cycle limit in which we approach the brane also
along $\mathbb{R}^2$. We can sum the series up to $O(1/k^2)$ for
the solutions of the homogeneous equation and up to $O(1/k^3)$ for
the particular solution of the non-homogeneous one:
\begin{eqnarray}
H^{(R)}(\mu,\, w)&=& H^{(R)}_0(k\,w)+\frac{1}{k}\,H^{(R)}_1(k\,w)+O(\frac{1}{k^2})\nonumber\\
H^{(I)}(\mu,\, w)&=& H^{(I)}_0(k\,w)+\frac{1}{k}\, H^{(I)}_1(k\,w)+O(\frac{1}{k^2})\nonumber\\
H^{(n-h)}(\mu,\, w)&=& \frac{1}{k}\,
H^{(n-h)}_0(k\,w)+\frac{1}{k^2}\,H^{(n-h)}_1(k\,w)+O(\frac{1}{k^3})
\label{unosuk}
\end{eqnarray}
The functions $H_l(k\,w)$ in the above expressions are obtained by
summing the terms proportional to $k^{n-l}\,w^n$ for the solutions
of the homogeneous equation and the terms proportional to
$k^{n-l-1}\,w^n$ in the series for $H^{(n-h)}(\mu,\, w)$. This
requires the computation as a function of n of the coefficient of
the $k^{n-l}$ or $k^{n-l-1}$ order term in the Hessenberg
determinants $F_n(\mu,0)$ and  $\tilde{F}_n(\mu)$. Such a
computation can be done but, for simplicity, we  focus only on the
cases $l=0,\,1$). It can be easily checked\footnote{For simplicity
we assume that the coefficients $c_0^R $ and $c_0^I$ do not depend
on $\mu$} that for $l=0$ we have:
\begin{eqnarray}
H^{(R)}_0(\xi)&=&c_0^R\, \sum_{n=0}^{\infty}\frac{\xi^n}{n!^2}=c_0^R\, I_0(2\,\sqrt{\xi})\nonumber\\
H^{(I)}_0(\xi)&=&c_0^I\, K_0(2\,\sqrt{\xi})\nonumber\\
H^{(n-h)}_0(\xi)&=&-h(\mu)\,I_0(2\,\sqrt{\xi})+h(\mu) \label{l0}
\end{eqnarray}
where we have set
\begin{equation}
\xi=|k\,w|=\mu^2\,(r^2-a^2)/8 <1 \label{fischio}
\end{equation}
Once again, eq.s (\ref{l0}) are a nice
confirmation of what we had predicted in the previous section by
the analysis of the differential equation in the near cycle limit.
>From our general power series solutions we retrieve the
result that the regular and irregular solutions of the homogeneous
equation in eqs. (\ref{l0}) have the same form as the massive
Green functions on $ \mathbb{R}^2$.  For a detailed comparison
with our previous results it suffices to observe that:
\begin{equation}
  \mu \, v = \mu \frac{\sqrt{r^4-a^4}}{2r} \simeq_{r\to a}
  \mu\,\frac{\sqrt{r^2-a^2}}{\sqrt{2}}= 2\,\sqrt{\xi}
\label{fiasco}
\end{equation}
Eq.s (\ref{l0}) allow also to identify the correct boundary values
of the constants appearing in the homogeneous solution. As
announced we must have:
\begin{equation}
  c_0^I = \frac{Q}{4 \pi a^2} \quad ; \quad c_0^R =0
\label{czeri}
\end{equation}
Indeed eq.s (\ref{l0}) show that near the cycle   the particular
integral $H^{(n-h)}$ goes to zero as foreseen in our general
discussion.

At first order in $1/k$ a particular integral of the non-homogeneous
equation, is simply $\widetilde{H}^{(n-h)}(\mu,\,w)\sim h(\mu)/k$ and
therefore the behavior of its Fourier transform $H ^{(n-h)}(\rho,\,r)$ in the
near-cycle limit is :
\begin{equation}
H^{(n-h)}(\rho,\,r)\sim
-\frac{1}{2\pi^2\,a^4}\,\log{\left(\frac{\rho}{\rho_0}\right)^2}
\label{hlog}
\end{equation}
where according to the discussion in section 3 we have assumed the holomorphic
function $\gamma$ to be a logarithm (see also the discussion in appendix C),
and $\rho_0$ is an integration constant acting as a regulator, not to be
confused with the enhan\c{c}on radius.
The next order terms ($l=1$) in eq.s
(\ref{unosuk}) are written in appendix \ref{diffeq}.

As far as the asymptotic behavior of the
solution at infinity $r\gg a$ is concerned we observe that the
first term in the r.h.s. of the equation for $H(\rho,\,r)$,
eq.(\ref{maindifEH}),
is just a resolved delta function for radially dependent test functions
$\Phi(r)$ in the $\mathrm{ALE_{EH}}$ space:
\begin{eqnarray}
\lim_{a\rightarrow 0}
\int_{ALE}\,\left[\frac{4}{\pi^2}\frac{a^4}{r^8}\right]\,\Phi(r)\,
d\Omega_{ALE}&=& \Phi(0)
\end{eqnarray}
Therefore for $r\gg a $ the differential equation
(\ref{maindifEH}) becomes homogeneous, namely its r.h.s. vanishes.
The corresponding equation in the Fourier transform $\widetilde{H}(\mu,\,r)$,
for $w\rightarrow -\infty$ has two independent solutions of which only
one is consistent with the boundary condition at infinity
(\ref{heun}), namely:
\begin{eqnarray}
\widetilde{H}(\mu,\,r)&=&c\,\frac{\mu}{r}\, K_1(\mu\,r) \simeq c
\times u_1(a,\,\mu,\,r)
\end{eqnarray}
where we have used the radial variable $r$ instead of $w\sim
-r^2/(2 a^2)$. The coefficient $c$ has already been fixed in terms
of the $3$-brane charge and the absence of the near cycle
regular solution, that is instead divergent at infinity
is confirmed.  As we expect in this limit, the behavior of
the warp factor $H$ is just the same as in the orbifold case
\cite{Bertolini:2000dk}, schematically:
\begin{equation} \left(\rho, r\gg a\right):\quad H(\rho,\,r)\sim
\frac{Q}{(\rho^2+r^2)^2}\left(1 + \mathrm{log~terms} \right)
\end{equation}
On the other hand, in the region near the cycle, combining the asymptotic
behaviors of the homogenoeus and inhomogeneous solutions, we find
{\large
\begin{equation}
\label{Hregcy}
\left(\matrix{ \rho< a\cr r\sim a }\right):\quad
H(\rho,\,r)\sim \frac{Q}{2a^2}\frac{1}{\rho^2+v^2} -
\frac{1}{2\pi^2\,a^4}\,K\,\log{\left(\frac{\rho}{\rho_0}\right)}^2
\end{equation}
}where $K$ is a (undetermined at the level of the present analysis) constant. The
corrections to the above behavior can be systematically deduced from the power
series expansion of the solution described in the present section.

Looking at eq.(\ref{Hregcy}), it is clear that there could be a
value of $r,\rho$ for which $H=0$, this being an indication for the
presence of a naked singularity of the repulson type \cite{kal}. This
singularity should be removed, somehow. In most non-confomal
$\mathcal{N}=2$ versions of the gauge/gravity correspondence this
singularity has been shown to be excised by the so-called
enhan\c{c}on mechanism \cite{enhanc}. This is the case, for instance, of
fractional branes on orbifolds, \cite{Bertolini:2000dk},
\cite{Anselmi:1994sm}. The value of the
enhan\c{c}on corresponds to $\rho=\rho_e$, the scale where the scalar
field $\gamma$ vanishes. This in general turns out to be the scale where the
dual gauge theory becomes strongly coupled and new light degrees of freedom
are expected to become relevant, both at the gauge theory (where instanton
effects become important) and at the supergravity level (where
tensionless strings occur). For all this analysis to work, it is
important that $\rho_e$ is bigger than the scale at which the repulson
occurs. In fact, when this is the case, the region where supergravity
is reliable, namely $\rho>\rho_e$, is free of any singularity. In
order to see if this happens also in our case and if the cut-off
$\rho_e$ has in fact the expected meaning, one should have a full
control on the world-volume action of the source, with all numerical
coefficient, normalization, etc... This we cannot do at the level of
the present pure supergravity analysis.

Our solution, however, differs from that of fractional branes on
singular space because of a better behaviour of the warp factor $H$ on the
plane $\mathbb{R}^2$ where it is non-singular, while the solution of the field
$\gamma$, which is responsible for the enhan\c{c}on mechanism,
has essentially the same structure. It is then natural to
believe that the nice properties described above should not drastically
change. Hence we consider our solution to be reliable, well-defined
and singularity free for $\rho>\rho_e$ (where the leading order
near cycle behaviour is well described by the first term in
eq. (\ref{Hregcy}) itself) while
we postpone a more precise analysis of this issue to future work.

\section{Discussion and Conclusions}
\label{conc}
In this paper we have constructed a new class of solutions of type
IIB supergravity describing 3-branes  whose transverse geometry
is of the type $M_6 = \mathbb{R}^2 \times ALE$ for an arbitrary
smooth $ALE$ manifold. We have shown that these solutions
preserve 8 supercharges and thus correspond to $\mathcal{N} = 2$ supersymmetry
in four dimensions. They are characterized by constant dilaton
and axion field, a 3-brane metric
\begin{equation}
ds^2 = H^{-\frac{1}{2}}\left (-\eta_{\mu\nu}dx^\mu\,dx^\nu \right )+
H^{\frac{1}{2}}dzd\bar{z}+H^{\frac{1}{2}}
ds^2_{ALE},
\end{equation}
a self dual five-form
\begin{equation}
F_{[5]}^{RR} =  d \left( H^{-1} \, \Omega_{\mathbb{R}^{1,3}} \right) +
\star  d \left( H^{-1} \, \Omega_{\mathbb{R}^{1,3}} \right),
\end{equation}
and a complex three-form, valued only in the transverse space,
whose form is uniquely fixed by supersymmetry:
\begin{equation}
  {\mathcal{H}}_+ =  2 \,d \gamma_I (z)\, \wedge \,
  \omega^I \label{tref}
\end{equation}
($\gamma_I (z)$ being analytic functions on $\mathbb{R}^2$
and  $\omega_I$ harmonic square-integrable anti-self-dual forms
on $ALE$).
The whole set of type IIB equations of motion reduces to solving
the following PDE for the warp factor $H$
\begin{equation}
  \left( \square_{\mathbb{R}^2} + \square_{ALE} \right)  H =
  - \partial_z \gamma_I \,  \partial_{\bar z}
  {\bar \gamma}_J \, \Delta^{IJ}(y) \label{repeatmain}
\end{equation}
This equation is quite more involved compared with the analogous
equations for the warp factor arising from $\mathcal{N}=1$
supersymmetry which are of the ODE type due to the conical nature
of the transverse space. We have then proceeded in studying eq.
(\ref{repeatmain}) in the simplest instance of an ALE manifold,
the Eguchi-Hanson space, whose properties were reviewed in
section \ref{Ehcase} and appendix \ref{EH}. In the case in which
the (here unique) analytic function $\gamma(z)$ possesses radial
symmetry, the equation for the warp factor took the form
\begin{equation}
\frac{1}{\rho}\partial_\rho
\rho\partial_\rho\,H+\frac{1}{r^3}\partial_r\left(r^3\,g(r)^2\,\partial_r
H\right) = - \frac{2\, a^4}{\pi^2\,r^8}\,|\partial\gamma(z)|^2
\label{repeatEH},
\end{equation}
and, after transforming to the variable $\mu$ conjugate to $\rho$,
reduces to a particular type of confluent Heun equation, whose
solution by power series can be determined with arbitrary
accuracy. Particularly interesting is the behavior of $H(\rho, r)$ near
the cycle $r=a$. It is still necessary to invoke an enhan\c{c}on-like
mechanism \cite{enhanc} to cut-off the region $\rho < \rho_e$
where supergravity is
not reliable. However, contrary to the orbifold case, the warp factor
$H(\rho, r)$ is regular as $r \to a$ for $\rho > \rho_e$, due to the
fact that the contribution of the three-form to the source term has
been regularized. In other words, wherever supergravity is
applicable, smoothing out  the transverse geometry does regularize
the solution for the warp factor.
The above result can be understood by studying the limit of
eq. (\ref{repeatEH})
in the presence of an auxiliary source term
\begin{equation}
\mbox{source} \propto - Q \, \frac{\delta(r-a)}{r^3}\delta^2(z)
\label{pasqua}
\end{equation}
on the r.h.s. introduced to fix  the boundary conditions near the
cycle. In the region $r \to \infty$ the Laplacian reduces to the flat
Laplacian on $\mathbb{R}^6$ and $Q$ is the total charge observed at
infinity (provided the complex 3-from is vanishing). In the opposite limit,
$r \to a$, the Laplacian reduces to the flat Laplacian on
$\mathbb{R}^4 \times S^2$ (the $S^2$ dependence of $H$ is trivial) and
the sources are clearly regular as long as $\rho \not= 0$, thus so is
the solution. By computing the density of charge $q$ on the two-cycle
one finds that it is consistent with the interpretation that the
three-brane charge is uniformly smeared over the cycle ($q=Q/\pi a^2$,
being $\pi a^2$ the area of the cycle), thus each element $d\Sigma$
of the surface area of the cycle contributes $q \, d\Sigma$ to the
charge at infinity.
The simple form of the Laplacian near the two-cycle, and the fact
that we have control over the boundary conditions, allows us to
write down the asymptotic form of the metric in that
region. Neglecting the contribution of the
three-form compared with the source term for the five-form
\footnote{The argument below becomes exact in the limit where the
three-form charge vanishes and it is applicable as long as the $\rho_e$ cut-off
is not too small.}
we find
\begin{equation}
\left( \frac{1}{\rho}\partial_\rho \rho\partial_\rho +
\frac{1}{v}\partial_v v\partial_v \right) H \propto
- q \delta(v^2)\delta(\rho^2)
\label{aproH}
\end{equation}
which is nothing but the Laplace equation for a point charge in four
dimensions written in ``bi-polar'' coordinates.
Introducing the variable $R = \sqrt{(\rho^2 + v^2)}$, denoting
the radial coordinate of $\mathbb{R}^4$ in units of $q$ we find
that $H \sim q/R^2$, yielding
\begin{equation}
ds^2 \sim \frac{R}{\sqrt{q}}(-\eta_{\mu\nu}dx^\mu dx^\nu) +
 \frac{\sqrt{q}}{R} dR^2+
\sqrt{q} R ds^2_{\mathbf{S}^3} + \frac{\sqrt{q}}{R}
\frac{a^2}{4} ds^2_{\mathbf{S}^2}
\end{equation}
where $ds^2_{\mathbf{S}^2}$ is the metric of the two cycle of
Eguchi-Hanson and $ds^2_{\mathbf{S}^3}$ is the metric of the
three-sphere at fixed $R$ in $\mathbb{R}^4$. An interesting point
is that the above result holds even in the absence of the
three-form flux and shows that conformal invariance of the dual
theory is always broken because the metric in $R$ and $x^\mu$ is
no longer of anti de Sitter type. Obviously we cannot take the
limit $R \to 0$, because this would imply going to large
curvatures, and in particular crossing the enhan\c{c}on radius.
However, the result suggests that a possible interpretation for
the parameter $a$ in the Eguchi-Hanson metric is that of a Fayet
Iliopoulos term, breaking conformal invariance in the infrared, in
accord with previous work on the subject.

The situation can be summarized as follows. In the absence of flux the
exact supersymmetric $3$-brane solution that we have found interpolates
between a standard ten dimensional $3$-brane solution at the singularity of
the metric cone on $S^5$, i.e. the standard $ \mathbb{R}^6$ manifold, and
a $3$-brane solution of an effective $8$-dimensional supergravity. As explained
in the literature \cite{popino1} (for a review see \cite{parilez}) we
can always consider sphere reductions of all supergravity theories and in
particular an $S^2$-reduction of type IIB supergravity.
This yields an effective
$8$-dimensional supergravity that has $3$-brane solutions.
In this case however,
there is a coupling to an effective dilaton  that emerges as the conformal
factor of the metric in the dimensional reduction. Hence the $3$-brane solution
of the $8$-dimensional supergravity is no longer conformal and we can follow
the prescription of Townsend et al \cite{Boonstra:1999mp} by making the transition
to a {\it dual frame} where the metric factorizes into the product of a
$3$-sphere
metric times a domain wall solution of an effective five-dimensional
supergravity
theory.
\par
Although the main thrust of this paper has been on the supergravity solution,
there are various properties of the dual gauge theory that can be described
to complete the analysis.
\par
We have seen that supergravity alone is not sufficient to
determine $\gamma(z)$ or the boundary conditions for the warp factor, while
this is easy in the case of the orbifold (by using for
example boundary state techniques) or on the conifold (in which one knows
a priori the structure of the source term, despite one does not know the
world--volume action).
This arbitrariness implies that our solution describes various
deformation or various vacua of ${\cal N}=2$ theories.
\par
Let us start by taking the orbifold limit and $\gamma(z) = const.$
What we obtain is clearly (N) regular D3-branes at the orbifold
point. The constituent branes could move away from the orbifold point
in all six transverse directions, thus the dual gauge theory has the
relevant hyper-multiplets and it is the well known conformal version
of an ${\cal N}=2$ Yang Mills with product gauge group $\prod_{I=1}^k\,\mathrm{U(d_I\,N)}$ ($k$ being the number of
irrepses  of the discrete kleinian group $\Gamma \subset \mathrm{SU(2)}$ that characterizes the orbifold and $\mathrm{d_I}$ their dimension)
and hyper-multiplets in the bi--fundamental representation. The blow up
 constitutes no problem in this case, and
it is well understood in terms of Fayet-Iliopoulos terms that break conformal invariance but preserve
supersymmetry.
In the simplest case, namely for $\Gamma=\mathbb{Z}_2$, which is the one we  focused
on, there are three such terms. 
\par
On the other hand, in the orbifold limit ($a\rightarrow 0$), it is
known that we may
switch  on a three form flux described by $\gamma(z)\neq        
\mbox{const.}$ which preserves the ${\cal N}=2$ supersymmetry. In the
unbroken phase symmetry requires that we can only have $\gamma(z)\,=\,z^\ell$
$(\ell>0)$ or $\gamma(z)\,=\,\log{(z)}$ $(\ell=0)$. In the case of a small flux,
the  corresponding operator
on the  boundary CFT side can be seen as a perturbation with  conformal
weight $\Delta\,=\,\ell+4$. This
perturbation is irrelevant for $\ell>0$, not affecting therefore the conformal IR fixed
point, and  it is marginal for $\ell=0$, signaling the presence of
a small number of fractional branes together with the regular ones at the
orbifold point.\par
The original and perhaps surprising feature of our solution consists in
the fact
that it may be thought of as the simultaneous effect on the CFT describing
regular D3--branes at the orbifold point of the two kind of deformations  
discussed above: the {\it relevant} one related to the blow--up moduli and
the {\it marginal/irrelevant} one related to the non--trivial flux.
Indeed, as previously pointed out,  the solution
$\gamma(z) \sim \log{(z)}$ occurs in the orbifold case
 for the fractional D3-brane solution
where the need for a logarithm is dictated by the source term
coming from the world volume action \cite{Bertolini:2000dk}.
Yet we know that in that case the
orbifold {\it cannot} be blown up \cite{das} and this can also be understood
from the fact that the gauge theory dual to the fractional brane lacks
the fields necessary to construct the appropriate Fayet--Iliopoulos terms.
One might have deemed that this fact would rule out solutions
of this kind in type IIB supergravity: on the contrary, in this paper we
have shown that such solutions are indeed possible, although they might not be interpreted as
the resolution of fractional branes on orbifolds.
\par
For small fluxes the most obvious interpretation is that we are perturbing
the (already resolved) theory on the regular brane as discussed above, but
there is also the possibility that some backgrounds with logarithmic behavior
correspond to blowing up a fractional brane not at the orbifold point
(background $B_{NS}$ field flux through the cycle equal to
$1/2$ in string units) but at the singular point (with zero flux), where
the no-go theorems \cite{das} do not apply. This raises the interesting issue
of whether our solution can,
in some sense, be seen as interpolating between the SCFT + FI theory with
product gauge group and the pure YM theory typical of a fractional brane.
\par
This is one instance in which supergravity alone is insufficient in
determining which is the dual gauge theory and we need string theory inputs as
discussed above.
This is not too surprising in the case of ${\cal N}=2$ theories since,
strictly speaking, supergravity alone is never sufficient to determine a
dual theory \'a la Maldacena, (for instance we cannot take the near horizon
limit without invoking the enhan\c{c}on mechanism).
\par
In the future, we plan to return to these issues by making a careful
study of the boundary action and trying to test if the above supergravity
picture can be confirmed by a microscopic analysis.
It would also be interesting to make connection with very recent
work on D-instantons \cite{fu} and wrapped fivebranes \cite{dario}
and \cite{zaffa}.
\vskip 1.cm
\noindent
{\large {\bf Acknowledgments}}
\vskip 0.5cm
\noindent
We wish to thank P. Di Vecchia, M. Kruczenski, A. Lerda, R. Russo for
discussions and very specially M. Bill\'o who has given essential and enlightening
advice throughout the whole development of this research project.
G.F. wishes to thank the organizers and the participants of the
Superstring Theory Workshop in Heraklion (particularly C. Bachas,
G.G. Gibbons, E. Kiritsis, U. Lindstr\"om, R. Russo and J. Sonnenschein) for
the many discussions, I. Klebanov for email correspondence, D.
Martelli, B.E.W. Nilsson and D. Tsimpis for discussions regarding
related supergravity solutions and U. Gran for software help. M.B. acknowledges
support by INFN.
\appendix

\section{Notations and Conventions}
\label{notazie} The supersymmetric formulation of type IIB
supergravity we rely on is that of Castellani and Pesando
\cite{igorleo} that uses the rheonomy approach
\cite{castdauriafre}. Hence, as it is customary in all the
rheonomy constructions, the adopted signature of the space-time metric
$d\tau^2$ used to set the conventions for the gamma matrices is the
mostly minus signature:
\begin{equation}
  \eta_{AB}= \mbox{diag} \left\{ + ,\underbrace{
  -,\dots,-}_{\mbox{9 times}} \right \}
\label{mostmin}
\end{equation}
However, in presenting the solution for the metric we
used $ds^2 = -d\tau^2$ to conform with more recent usage.

The index conventions are the following ones:
\begin{eqnarray}
A,B,\dots & = & 0,\cdots,9 \quad \mbox{Flat indices in $D=10$} \nonumber\\
a,b,\dots & = & 0,\cdots,3 \quad \mbox{Flat indices on
$\mathbb{R}^{1,3}$}
\nonumber\\
\alpha, \beta, \dots & = & 4,5 \quad \mbox{Flat indices
on $\mathbb{R}^2$} \nonumber\\
t,s,\dots & = & 6,7,8,9 \quad \mbox{Flat indices on ALE}\nonumber\\
(i,j,\dots & = & 4,\dots,9 \quad \mbox{Flat indices on
$\mathbb{R}^2 \times \mbox{ALE}$})\nonumber\\
&& \nonumber \\
M,N,\dots & = &  \quad \mbox{Curved indices in $D=10$} \nonumber\\
\mu,\nu,\dots & = &  \quad \mbox{Curved indices on $\mathbb{R}^{1,3}$} \nonumber\\
z, \bar z & = &  \quad \mbox{Curved indices on $\mathbb{R}^2$} \nonumber\\
\tau, \sigma, \dots & = & \quad \mbox{Curved indices on ALE} \nonumber\\
&& \nonumber \\
I,J, \dots & = & 1,2,\dots, k \quad \mbox{indices labeling ALE harmonic
self dual 2-forms}\nonumber\\
\label{indeconv}
\end{eqnarray}
The complete vielbein of 10-dimensional space-time are denoted
$V^{A}$, those of the $\mathbb{R}^2 \times  \mathrm{ALE}$ manifold
without warping factor are denoted $e^i$. The convention for
constructing the dual of an $n$-form $\omega$ in 10-dimensions
is the following:
\begin{equation}
  \omega= \omega_{{A_1\dots A_n}} \, V^{{A_1}} \wedge \dots \wedge
  V^{{A_n}} \quad \Leftrightarrow \quad \star \omega = \frac{1}{(10-n)!} \,
  \epsilon_{{A_1 \dots A_{10-n} B_1
  \dots B_n}} \omega^{{B_1\dots
  B_n}} \, V^{{A_1}} \wedge \dots \wedge V^{{A_{10-n}}}
\label{hodgedual}
\end{equation}
Note that we also use $n$-form components with \emph{strength
one}: $\omega= \omega_{{A_1\dots A_n}} \,
V^{{A_1}} \wedge \dots \wedge
  V^{{A_n}}$ and not with strength $n!$ as it would
  be the case if we were to write $\omega= \frac {1}{n !}\omega_{{A_1\dots A_n}} \,
  V^{{A_1}} \wedge \dots \wedge V^{{A_n}}$.
When it is more appropriate to use curved rather than flat indices
then the convention for Hodge duality is summarized by the
formula:
\begin{equation}
  \star\left(dx^{M_1}\wedge\dots
dx^{M_n}\right)=\frac{\sqrt{-{\rm det}(g)}}{(10-n)!}g^{M_1 N_1}\dots
g^{M_n N_n}\epsilon_{L_1\dots
 L_{10-n}N_1\dots N_n}\,dx^{L_1}\wedge\dots dx^{L_{10-n}}
\label{hodgecurvo}
\end{equation}
Applying the rule (\ref{hodgecurvo}) to the case of the specific
$10$-dimensional metric of our ansatz (\ref{ansazzo}) we obtain
the following useful formulae for Hodge duals:
\begin{eqnarray}
\nonumber\\
\star\left(dz\wedge \Omega_{\mathbb{R}^{1,3}}\right)&=& -{\rm i}H^2\,
dz\wedge \Omega_{ALE}\nonumber\\
\star\left(dz\wedge \Omega_{ALE}\right)&=&  {\rm i}H^{-2}\,dz\wedge
\Omega_{\mathbb{R}^{1,3}}\nonumber\\
\star\left(dy^\tau\wedge\Omega_{\mathbb{R}^{1,3}}\right)&=&-\frac{{\rm
i}}{2} H^2\,(\hat{\star}dy^\tau)\wedge dz\wedge
d\bar{z}= -H^2\,(\hat{\star}dy^\tau)\wedge dx^4\wedge dx^5\nonumber\\
\star\left(dz\wedge \omega^I\right)&=&-{\rm i} H^{-1}\,dz
\wedge \omega^I\wedge \Omega_{\mathbb{R}^{1,3}}\nonumber\\
\Omega_{\mathbb{R}^{1,3}}&=& dx^0\wedge\dots dx^3\nonumber\\
\Omega_{ALE}&=&\sqrt{{\rm det}\,(g^{ALE})}\,dy^1\wedge\dots dy^4
\label{hodge}
\end{eqnarray}
where $\hat{\star}$ denotes the Hodge dual in the ALE space. The
Laplacians on $\mathbb{R}^2$ and the ALE space are:
\begin{eqnarray}
\square_{\mathbb{R}^2}&=&\left(\frac{\partial}{\partial
x^4}\right)^2+\left(\frac{\partial}{\partial
x^5}\right)^2\nonumber\\
\square_{ALE}\,(\cdot)&=&-d_{(ALE)}\hat{\star}\,d_{(ALE)}\,(\cdot)=
\nonumber\\&&\frac{1}{\sqrt{{\rm
det}\,(g^{ALE})}}\partial_\tau\left[\sqrt{{\rm det}\,(g^{ALE})}\,
(g^{ALE})^{\tau\sigma}\partial_\sigma\,(\cdot)\right]
\end{eqnarray}
 \noindent For
the gamma matrices our conventions are as follows:
\begin{eqnarray}
  \left\{ \Gamma^{A} \, , \, \Gamma^{B} \right\} & = & \eta^{AB}
\nonumber\\
\Gamma^{A}& =& \left\{ \begin{array}{rcl}
  \Gamma^a & = & \gamma^a \,\otimes {\bf 1} \\
  \Gamma^i & = & \gamma_5 \, \otimes \, \tau^i \,
\end{array} \right.
\label{grossgamma}
\end{eqnarray}
where:
\begin{equation}
\begin{array}{rclcrcl}
\left\{ \gamma^a \, , \, \gamma^b\right\}  & = & 2\, \eta^{ab}  &
; & \gamma_5 & =& - {\rm i} \gamma^0 \gamma^1 \gamma^2  \gamma^3
\\
\left\{ \tau^i \, , \, \tau^j\right\}  & = & 2\, \eta^{ij} = - 2
\, \delta^{ij} & ; & \tau_\bullet &=& - \tau^4 \tau^5  \tau^6
\tau^7  \tau^8 \tau^9\
\end{array}
\label{taumatte}
\end{equation}
are the gamma matrices in Lorentzian four space and on the six
dimensional manifold $M_6$ respectively. The matrices $\gamma_5$
that anticommutes with all $\gamma^a$ and $\tau_\bullet$ that
anticommutes with all $\tau^i$ are respectively hermitian and
antisymmetric and we have:
\begin{equation}
  \gamma_5^2 ={\bf 1} \quad ; \quad \tau_\bullet^2 = -{\bf 1}
\label{gammezinque}
\end{equation}
Since the signature of $M_6$ is completely negative the
$\tau^i$ matrices that realize the associated Clifford algebra can
be taken to be all real and antisymmetric. They can be further
decomposed with respect to the submanifolds $ \mathbb{R}^2$ and
$\mathrm{ALE}$ as it follows:
\begin{equation}
  \tau^i =\left\{ \begin{array}{rcl}
    \tau^\alpha & = & {\rm i} \sigma^{\alpha} \, \otimes \, {\bf 1} \\
    \tau_u & = & {\rm i} \sigma^3 \, \otimes \, \overline{\gamma} _u\
  \end{array} \right.
\label{tausplitte}
\end{equation}
where $\sigma^\alpha = \sigma^{1,2}$ and $\sigma^3$ are the
standard Pauli matrices while:
\begin{equation}
 \left\{  \overline{\gamma} _u \, , \,  \overline{\gamma} _v \right\}
 = \delta_{uv}
\label{euclidgam}
\end{equation}
are $4 \times 4 $ hermitian matrices forming an Euclidean
realization of the four-dimensional Clifford algebra.

\section{The Eguchi-Hanson space}
\label{EH}
In the present appendix we state our conventions for
the description of the Eguchi-Hanson space. The metric is defined
in eq. (\ref{eguchi}) of section \ref{Ehcase}, where the
$1$-forms $\sigma^i=\left\{ \sigma^x \, \sigma^y \, \sigma^z
\right \}$ are an explicit realization of the Maurer Cartan forms
of the $\mathrm{SU(2)}$-Lie algebra:
\begin{equation}
  d\sigma^i = \epsilon^{ijk} \sigma^j \wedge \sigma^k
\label{su2maucart}
\end{equation}
and their expression is:
\begin{eqnarray}
\sigma_x&=&-\frac{1}{2}\left(\cos{\psi}\,d\theta +\sin{\theta}\,\sin{\psi}\,d\phi\right)\nonumber\\
\sigma_y&=&\frac{1}{2}\left(\sin{\psi}\,d\theta -\sin{\theta}\,\cos{\psi}\,d\phi\right)\nonumber\\
\sigma_z&=&-\frac{1}{2}\left(d\psi+\cos{\theta}\, d\phi\right)
\label{sigmai}
\end{eqnarray}
Let us now consider the properties of the homology $2$-cycle
which, in this case, is unique. In the coordinates
$(\psi,r,\theta,\phi)$ the cycle is spanned by $(\theta,\phi)$
setting $r=a$. At $r=a$  the Eguchi-Hanson metric has a
Bolt singularity which is a coordinate singularity given that
$\psi$ has period $2\pi$. The limit $r\rightarrow a$ is therefore
regular for the solution, as can be verified by changing
coordinates from r to $u=r^2\, g(r)^2$ and taking $u\rightarrow
0$. Restricting the metric to $\theta,\phi$ with $r=a$ yields:
\begin{eqnarray}
ds^2_{\Sigma}&=&
\frac{a^2}{4}\left(d\theta^2+\sin{\theta}^2\,d\phi^2\right)
\end{eqnarray}
that describes the exceptional cycle $\Sigma$ as a sphere of
radius $a/2$. \par Next let us write the explicit form of the
harmonic anti self-dual $2$-form $\omega$ fulfilling properties
(\ref{omega}). Since $\sigma_z$ is not globally defined on
$\Sigma$ the form:
\begin{eqnarray}
\omega &=&\frac{a^2}{2\,\pi}\,d\left(\frac{\sigma_z}{r^2}\right)=
 \frac{a^2}{2\pi r^3}\, dr\wedge d\psi+ \frac{a^2}{2\pi r^3}\, \cos{\theta}\,dr\wedge d\phi+\frac{a^2}{4\pi r^2}\,
\sin{\theta}\,d\theta\wedge d\phi
\end{eqnarray}
is closed but not exact and constitutes the only normalizable
representative of the second cohomology group of the space. It can
be checked that: $\int_{\Sigma}\,\omega=1$,
$\int_{ALE}\,\omega\wedge \omega=-1/2=-{\cal C}^{-1}$ and that
$\star\omega=-\omega$. To check the last property we fix our
conventions for the Levi-Civita tensor
$\epsilon_{\psi\,r\,\theta\,\phi}=1$ and give the following useful
formulae:
\begin{eqnarray}
\star\left(d\theta\wedge d\phi\right)&=&
-\frac{2\cos{\theta}}{r\sin{\theta}}\, dr\wedge d\phi-\frac{2}
{r\sin{\theta}} \, dr\wedge d\psi\nonumber\\
 \star\left(dr\wedge d\psi\right)&=& -\frac{r}{2\sin{\theta}}\left(1-\left(\frac{a}{r}
\right)^4\, \cos{\theta}^2\right)\, d\theta\wedge
d\phi-\frac{r\cos{\theta}}{2
\sin{\theta}}\, g(r)^2\, d\theta\wedge d\psi\nonumber\\
 \star\left(dr\wedge d\phi\right)&=& \frac{r}{2\sin{\theta}}\, g(r)^2\, d\theta\wedge d\psi+\frac{r\cos{\theta}}{2
\sin{\theta}}\, g(r)^2\, d\theta\wedge d\phi
\end{eqnarray}
where we have used the expression for the determinant of the
Eguchi-Hanson metric:
\begin{equation}
  {\rm det}\, (g^{ALE})=(r^3\,
\sin{\theta}/8)^2
\label{determino}
\end{equation}
Finally the expression (\ref{Delta}) for
the function $\Delta(y)$ is easily derived from the following
wedge product:
\begin{eqnarray}
\omega\wedge \omega &=& -\Delta(y)\,\Omega_{ALE} \quad ; \quad
\Omega_{ALE}=\frac{r^3}{8}\sin{\theta}\,d\psi\wedge dr\wedge
d\theta\wedge d\phi
\end{eqnarray}

\section{Analysis of the equation for H in the Eguchi-Hanson space}
\label{diffeq}
The present appendix is to put on a more rigorous
mathematical ground the results on the study of the warp factor $H$
discussed in section \ref{Ehcase} and to fix the relevant
conventions. The general definition of the Fourier transform
$\widetilde{H}(\vec{p},y)$ ($\vec p$ here is the momentum vector along
$\mathbb{R}^2$) is
\begin{eqnarray}
 \widetilde{H}(\vec{p}, y)& = &
\frac{{\rm i}}{2}\, \int \, \frac { dz\wedge d{\bar z}
 }{(2\pi)^2}\, \exp\left [ {\rm i} (p {\bar z} + {\bar p} z)/2 \right ]
 \, \left( H(z,{\bar z} , y ) -1\right) \nonumber\\
 &\equiv &\widetilde{H}(p,\bar p,y)
\label{genfourier}
\end{eqnarray}
As already noticed in section 4, in eq.(\ref{genfourier}) the two dimensional
momentum vector $\vec{p}$ was substituted by a complex number $p$ whose norm
we name $\mu$:
\begin{equation}
  p = p_4 + {\rm i} p_5 \quad ; \quad \sqrt{|p|^2} \equiv \mu
\label{pvector}
\end{equation}
while $\rho$ denotes the radius in $\mathbb{R}^2$.
In section \ref{Ehcase} we dealt with the case in which $|\partial
\gamma(z) |^2$ depended only on $\rho$. In this situation we could
expect the integral $H(z,\bar z, y)$ to depend only on
$\rho$ and $r$ since the source term and the coefficients of its
differential equation would depend only on these radial
coordinates. Since $\gamma(z)$ is required to be analytic, this
 occurs when $\gamma(z)$ is either a logarithm or a simple power of $z$:
\begin{equation}
  \gamma(z) \sim \cases {\log (z) \cr
  \frac{z^\alpha}{\alpha}\cr}
\label{specgamma}
\end{equation}
Defining the Fourier transform of $|\partial \gamma(z)
|^2$:
\begin{eqnarray}
j(p,\bar p) & \equiv & \frac{{\rm i}}{2} \, \int \, \frac {
dz\wedge d{\bar z}
 }{(2\pi)^2}\, \exp\left [ {\rm i} (p {\bar z} + {\bar p} z)/2 \right ] \,
\vert \partial \gamma(z) \vert^2
\label{gracida}
\end{eqnarray}
$\zeta$ being the angular variable on $\mathbb{R}^2$. In the two
special cases (\ref{specgamma}) the Fourier transformed source
term $j$ is a function only of $\mu$ since we can write:
\begin{eqnarray}
  j(\mu) &=& \frac 1 {(2 \pi)^2} \, \int_0^{\infty} \rho d\rho \
\int_0^{2\pi} \,d \zeta \, \exp \left[ {\rm i } \,\rho \, \mu \,
\cos \zeta \right ] \, \vert \partial
\gamma\vert^2(\rho)=\frac{1}{2\pi}\int_0^{\infty} \rho d\rho \,
J_0(\mu\,\rho) \, \vert \partial \gamma\vert^2(\rho)\nonumber\\
&&
\end{eqnarray}
which implies\footnote{The Fourier-Bessel transform of $|\partial
\gamma|^2$ in the logarithmic case is a divergent constant,
independent from $\mu$. Formally we name it $j(log)$ to remind
ourselves what it is. At the end, one takes the
inverse Bessel-Fourier transform and $j(log)$ is restored to
$1/\rho^2$ }
\begin{equation}
 j(\mu) = \left \{
 \begin{array}{rrcl}
 \lim_{\alpha\to 0} & \frac{1}{2\pi}\int_0^{\infty} \rho
 d\rho \, J_0(\mu\rho) \,\frac {1}{\rho^{-2\alpha+2}} &\equiv& j(log) \to \infty \\
   \null & \frac{1}{2\pi}\int_0^{\infty} \rho
 d\rho \, J_0(\mu\rho) \,\frac {1}{\rho^{-2\alpha+2}} & = & \left( \frac{4^{\alpha -1}\,
      {\Gamma}(\alpha )}{\pi \,
      {\left( {\mu }^2 \right) }^{\alpha }\,
      {\Gamma}(1-\alpha )} \right)  \end{array}\right.
\end{equation}
and where we used the following integral representation of the
Bessel $J_0$ function:
\begin{equation}
\int_0^{2\pi} \,d \zeta\, \exp \left[ {\rm i } \,x \, \cos \zeta
\right ]= 2\pi J_0(x) \label{jointeg}
\end{equation}
Under the hypothesis (\ref{specgamma}) we can denote the  unknown
function $H$ by $H(\rho,\,r)$ and its Fourier transform
$\widetilde{H}(\mu,\,r)$ will be given by eq. (\ref{hspecial}). In
all other cases $\widetilde{H}=\widetilde{H}(p,\,\bar p,\,r)$.
\par
However, independently from the explicit form of the twisted field
$\gamma(z)$, after partial Fourier transform the original equation
(\ref{maindifEH}) reduces always to:
\begin{equation}
  L_r[a,\mu] \, \widetilde{H} = J(r,\,p,\,\bar{p})
\label{formalform2}
\end{equation}
which generalizes eq. (\ref{formalform}) to a more general choice
of $\gamma(z)$ and  where the differential operator $L_r[a,\mu]$
was defined in eq. (\ref{diffoperat}). In eq. (\ref{formalform2})
we have defined the complete source function:
\begin{equation}
  J(r,\theta,\mu)  \, \equiv \, -\frac{2 a^2}{\pi^2 \,
  r^2}\, j(p,\,\bar{p})+S(\mu,\,r)
\label{source2}
\end{equation}
 The Wronskian for a basis of two independent solutions
$u_1(a,\,\mu,\,r)$ and $u_2(a,\,\mu,\,r)$ of the homogeneous
equation (\ref{hom}) associated with eq. (\ref{formalform2}) is:
\begin{equation}
  W(u_1,u_2) = \mbox{det} \left( \begin{array}{cc}
    u_1(r) & u_2(r) \\
    u^\prime_1(r) & u^\prime_2(r) \
  \end{array}\right) = \mbox{const} \, \times \, \frac{1}{p(r)} = \mbox{const} \, \times \, \frac{1}{r^3 \, g(r)^2}
\label{wronskio}
\end{equation}
where the overall constant can be adjusted by choosing the overall
normalization of the solutions $u_1$ and $u_2$.
\par
Focusing on the case in which the source term $J$ depends
only on $\rho,\,r$, the non-homogeneous equation
(\ref{formalform2}) can be solved, according to eq.
(\ref{gensolgrun}),  in terms of the radial Green function
$G\left( a,\mu\vert r,r^\prime \right)$ defined by eq.
(\ref{grunadefi}) plus the the boundary
conditions to be imposed on the solutions $\widetilde{H}$.
Following  the general theory \cite{dennery} the Green function is
constructed in terms of a basis $u_1,u_2$ of solutions of the
homogeneous equation as follows:
\begin{equation}
  G(r,r^\prime) = \cases{ u_< = c_1(r^\prime) \, u_1(r) + c_2
  (r^\prime) \, u_2(r) \quad \mbox{for} \quad r < r^\prime \cr
  u_> = d_1(r^\prime) \, u_1(r) + d_2
  (r^\prime) \, u_2(r) \quad \mbox{for} \quad r > r^\prime \cr}
\label{grunapiumeno}
\end{equation}
where the coefficients $c_i,\,d_i$ become functions of the second
variable and are determined by:
\begin{enumerate}
  \item the continuity of $G$ at $r=r^\prime$:
\begin{equation}
  u_<(r^\prime) = u_>(r^\prime)
\label{continuity}
\end{equation}
  \item the discontinuity of $1/w(r)$ in the derivative   $\partial
  G$ at $r=r^\prime$:
\begin{equation}
  \lim_{\epsilon \to 0} \, \left\{ \frac{d}{dr}G\left (r,r^\prime\right)
  \vert_{r=r^\prime+\epsilon} - \frac{d}{dr}G\left (r,r^\prime\right)
  \vert_{r=r^\prime-\epsilon}\right\} =  \frac{1}{r^3 g(r)^2}
\label{discontinuo}
\end{equation}
\item The symmetry of the real Green function in the exchange of its
arguments:
\begin{equation}
  G(r,r^\prime) = G(r^\prime , r)
\label{grumsymm}
\end{equation}
  \item The boundary conditions at the end-points of the interval
  $[a,\infty]$ that we want to impose on the function
  $\widetilde{H}(r)$:
\begin{equation}
  \lim_{r\to a} \widetilde{H} = h_a \quad ; \quad \lim_{r\to \infty}
  \widetilde{H}= h_\infty
\label{boundario}
\end{equation}
The same conditions have to be imposed on the Green
function.
\end{enumerate}
While the first three items in the above list  are mathematical
requirements, the fourth encodes the definition of the physical
system we want to describe. For this reason we have to choose the
boundary conditions by careful comparison with the flat case which
is our only guide at this stage. Furthermore, since the solutions
$u_1,u_2$ will be shown to have singular behaviors at the end
points of the interval, eq.(\ref{boundario}) has to be understood
in the sense of asymptotic regimes.
\par
As shown in \cite{dennery}, the first two conditions become a
linear system of equations for the $r^\prime$-dependent
parameters:
\begin{equation}
  c=c_1-d_1 \quad ; \quad d=c_2-d_2
\label{cdparameters}
\end{equation}
with the general solution:
\begin{equation}
  c(r^\prime) =\frac{u_2(r^\prime)}{(r^\prime)^3 \, g(r^\prime)
W(r^\prime)}\quad ; \quad d(r^\prime)
=-\frac{u_1(r^\prime)}{(r^\prime)^3 \, g(r^\prime) W(r^\prime)}
\label{solvcdparam}
\end{equation}
Recalling eq.(\ref{wronskio}) if we normalize $u_1$ and $u_2$ in
such a way that the constant appearing in the Wronskian is fixed
to $2$:
\begin{equation}
  W(u_1,u_2) = \frac{2}{r^3 \, g(r)^2}
\label{fissonorma}
\end{equation}
then the Green function takes the form \cite{dennery}:
\begin{equation}
  G(r,r^\prime)=G_s(r,r^\prime) + d_1(r^\prime) u_1(r) +d_2(r^\prime)
  \, u_2(r)
\label{GandGs}
\end{equation}
where:
\begin{equation}
  G_s(r,r^\prime)=\left \{ \begin{array}{cl}
  \ft 1 2 \left( u_2(r^\prime) \, u_1(r) - u_1(r^\prime) \,
  u_2(r)\right) & \mbox{for}\quad r < r^\prime \\
  0 & \mbox{for}\quad r > r^\prime \ \end{array} \right.
\label{Gsonly}
\end{equation}
To fix the last bit of information, namely $d_1$ and $d_2$, we
have to impose a choice of boundary conditions. The analysis of
sections  \ref{asymptsect} and \ref{d3branacargo} allows to
determine the required near cycle behavior of the integral
$\widetilde{H}$ functions and consequently both the expressions
for the source term $S(\rho,\,r)$ (\ref{risorgo}) and for the two
functions $d_1$ and $d_2$:
\begin{equation}
  d_1(r^\prime) = -\ft 12 \, u_2(r^\prime) \quad ; \quad d_2(r^\prime) =0
  \label{trulagruna2}
\end{equation}
which determine in turn the form of the Green function, given in
eq. (\ref{trulagruna}).

In this final part of the
appendix we derive some of the main results of the analysis
carried out in section \ref{heunsoft}, besides describing some
general feature of eq. (\ref{heun}).\par
 Equation (\ref{heun}) is
characterized by two regular singularities of the homogeneous part
at the points $w=0,\, 1$. The regular singularity $w=0$
corresponds to the position of the exceptional cycle $r=a$ while
the irregular singularity at $w=-\infty$ corresponds to the large
distance limit $r\rightarrow \infty$. The homogeneous equation
associated with (\ref{heun}) can be framed within the class of
equations known as confluent Heun differential equations. A
confluent Heun equation may be written in its standard form as
follows:
\begin{eqnarray}
\frac{d^2\,
f}{dw^2}+\left(4\,p+\frac{\gamma}{w}+\frac{\delta}{w-1}\right)\,
\frac{d\,f}{dw}+\frac{(4\,p\,\alpha\, w-\sigma)}{w\,(w-1)}\, f&=&0
\label{heungen}
\end{eqnarray}
and  in general it has two kind of solutions: an {\it angular}
solution $f^{(a)}(p,\alpha,\gamma,\delta,\sigma;z)$ and a {\it
radial} solution $f^{(r)}(p,\alpha,\gamma,\delta,\sigma;z)$,
defined as solutions of suitable boundary value problems on the
intervals $]0,1[$ and $]-\infty,0[$ respectively. The homogeneous
part of eq. (\ref{heun}) is obtained from eq. (\ref{heungen}) for
$\gamma=\delta=1$, $\alpha=k/(2 p)+1$, $\sigma = k+2p$ in the
limit $p\rightarrow 0$. Although the literature on this confluent
limit of the Heun equation is rather limited, for our purposes we
 could restrict our analysis of the solution of (\ref{heun}) to the
neighborhood of the two physically relevant singular points,
namely $w=0$ and $w=-\infty$ (i.e. to a region close to the
exceptional cycle and to radial infinity) and discuss the
restrictions imposed on the solution in these two regions by the
boundary conditions on the interval $]-\infty,\,0[$. As far as the
singularity $w=0$ is concerned, in its neighborhood we can
write a formal power series expansion for an integral of the
homogeneous equation:
\begin{eqnarray}
H(\mu,\,w)&=& w^\sigma \sum_{n=0}^{\infty}\, c_n(\mu,\,\sigma)\,
w^n
\end{eqnarray}
the determination of the coefficients $c_n(\mu,\,\sigma)$ through
recurrence relations is a standard procedure in dealing with
differential equations with regular singularities.
 Substituting the above series in the homogeneous equation we
obtain:
\begin{eqnarray}
L_w\,H&=&\sum_{n=0}^{\infty}\,\left(c_n(\mu,\,\sigma)\,
f_0(\sigma+n)+c_{n-1}(\mu,\,\sigma)\,
f_{1}(\mu,\,\sigma+n-1)+\dots \right.
\nonumber\\
&&\left. + \, c_0(\mu,\,\sigma)\,
f_n(\mu,\,\sigma)\right)\, w^n=0\nonumber\\
\label{prerec}
\end{eqnarray}
where the coefficients $f_{0}(\sigma+m)$ and
$f_{n>0}(\mu,\,\sigma+m)$ are determined by the following
expansion:
\begin{eqnarray}
(\sigma+m)(\sigma+m-1)+(\sigma+m)\, P_1(w)+
P_2(\mu,\,w)&=&f_0(\sigma+m)+\sum_{n=1}^{\infty}\,f_n(\mu,\,\sigma+m)\, w^n\nonumber\\
f_0(\sigma+m)&=&(\sigma+m )^2\nonumber\\
f_1(\mu,\,\sigma+m)&=&k-\sigma-m\nonumber\\
f_{n>1}(\mu,\,\sigma+m)&=&-(k+\sigma+m)
\end{eqnarray}
Hence equation (\ref{prerec}) gives recursion relations for the
unknown coefficients $c_n(\mu,\,\sigma)$. The first of them reads
$c_0(\mu,\,\sigma)\, f_0(\sigma)=0$ and yields a non trivial
solution only if the indicial equation is fulfilled, namely if
$f_0(\sigma)=\sigma^2=0$. The coefficient $c_0(\mu,\,\sigma)$ is
thus a free parameter in terms of which all the other coefficients
are expressed. The solution to a generic order $n$ of the
recursion relations is represented in eq. (\ref{coefreg}) that we repeat here in a more explicit form:
\begin{eqnarray}
c_n(\mu,\,\sigma) &=&(-)^n\,
c_0(\mu)\,\frac{F_n(\mu,\,\sigma)}{\prod_{j=1}^{n}\,
f_0(\sigma+j)}=(-)^n\,
c_0(\mu)\,\frac{F_n(\mu,\,\sigma)}{\prod_{j=1}^{n}\,
(\sigma+j)^2}\nonumber\\
&&\nonumber\\
F_n(\mu,\,\sigma)&=&\left|\matrix{f_1(\mu,\,\sigma+n-1) &
f_2(\mu,\,\sigma+n-2) &\dots & f_{n-1}(\mu,\,\sigma+1)
&f_n(\mu,\,\sigma)\cr f_0(\sigma+n-1) & f_1(\mu,\,\sigma+n-2)
&\dots & f_{n-2}(\mu,\,\sigma+1) &f_{n-1}(\mu,\,\sigma)\cr 0 &
f_0(\sigma+n-2) &\dots & f_{n-3}(\mu,\,\sigma+1) &
f_{n-2}(\mu,\,\sigma)\cr \vdots &\vdots & &\vdots &\vdots\cr 0 & 0
& \dots & f_{0}(\sigma+1) &f_{1}(\mu,\,\sigma)} \right| \nonumber\\
\label{coefreg2}
\end{eqnarray}
Since
$\sigma =0$ is a double solution of the indicial equation, from
standard theory of differential equations it follows that the
homogeneous equation will have a regular solution
$H^{(R)}(\mu,\,r)$ represented by a power series expansion around
$w=0$ whose coefficients are $c_n^R(\mu)=c_n(\mu,\,\sigma=0)$, and
an irregular one $H^{(I)}(\mu,\,r)$, both these independent
solutions are defined in eq. (\ref{hir}).\par As far as the
particular solution $H^{(n-h)}(\mu,\,r)$ of the non-homogeneous
equation is concerned, we can write a formal power series
expansion as in eq. (\ref{particula}). Expanding the source term
on the right hand side of eq. (\ref{heun}) as in eq. (\ref{ss})
and substituting both eq.s (\ref{particula}) and (\ref{ss}) in eq.
(\ref{heun}) we obtain:
\begin{eqnarray}
&&L_w\,H^{(n-h)}-\frac{h(\mu) \,w}{(1-w)(1-2w)^3} =
c^{(n-h)}_0(\mu)\, f_0(0)+\nonumber\\ &&
\sum_{n=1}^{\infty}\,\left(c^{(n-h)}_n(\mu)\,
f_0(n)+c^{(n-h)}_{n-1}(\mu)\, f_{1}(\mu,\,n-1)+\dots
\right.\nonumber\\
&& \left.
+ \, c^{(n-h)}_0(\mu)\,f_n(\mu,\,0)-h(\mu)\, s_{n-1}\right)\, w^n=0\nonumber\\
\label{prerecih}
\end{eqnarray}
>From the above equation we may deduce the recursion relations for
$c^{(n-h)}_n(\mu)$ which are solved in eq. (\ref{coefih}).
Such a solution is repeated here in more explicit form:
\begin{eqnarray}
c_n^{(n-h)}(\mu)
&=&(-)^n\,\frac{\tilde{F}_n(\mu)}{\prod_{j=1}^{n}\,
f_0(j)}=(-)^n\,\frac{\tilde{F}_n(\mu)}{\prod_{j=1}^{n}\,
j^2}\nonumber\\
\tilde{F}_n(\mu)&=&\left|\matrix{f_1(\mu,\,n-1) & f_2(\mu,\,n-2)
&\dots & f_{n-1}(\mu,\,1) &-h(\mu)\,s_{n-1}\cr f_0(n-1) &
f_1(\mu,\,n-2) &\dots & f_{n-2}(\mu,\,1) &-h(\mu)\,s_{n-2}\cr 0 &
f_0(n-2) &\dots & f_{n-3}(\mu,\,1) & -h(\mu)\,s_{n-3}\cr \vdots
&\vdots & &\vdots &\vdots\cr 0 & 0 & \dots & f_{0}(1) &
-h(\mu)\,s_{0}} \right| \label{coefih2}
\end{eqnarray}
Finally let us consider the limit $\mu\gg 1/a$, $|w|\ll 1$ and
$|k\,w|<1$ in which the above series expansions can be summed
order by order in $1/k$ according to the expansions
(\ref{unosuk}). In eq.s (\ref{l0}) we gave the sum of the series
at the first non trivial order.\par For completeness we give below
the  sums at the next order in $1/k$:
\begin{eqnarray}
H^{(R)}_1(\xi)&=&c_0^R\, \sum_{n=0}^{\infty}\frac{\left( -2 + n
\right) \,\left( -n + n^2 \right) }{3}\frac{\xi^n}{n!^2}=\nonumber\\
&& c_0^R\, \frac{\xi^{3/2}}{3}\,I_3(2\,\sqrt{\xi})\nonumber\\
H^{(I)}_1(\xi)&=&c_0^I\,\left[2\,I_0(2\,{\sqrt{\xi }})\,\left( -1
+ \xi \,\log (\xi ) \right)  +
    {\sqrt{\xi }}\,\left( -\left( I_1(2\,{\sqrt{\xi }})\,
          \left( 6 + \left( 2 + \xi  \right) \,\log (\xi ) \right)  \right)
          +\right.\right.\nonumber\\&&\left.\left.
       \xi \,\left( 2\,K_3(2\,{\sqrt{\xi }}) +
          I_3(2\,{\sqrt{\xi }})\,\log (\xi ) \right)  \right)\right]\nonumber\\
H^{(n-h)}_1(\xi)&=&-\sum_{n=0}^{\infty}\left[s_0\,\frac{-3 + 2\,n
- 3\,n^2 + n^3}{3}-s_1\right]\frac{\xi^n}{n!^2}=
\nonumber\\
&&-\frac{h(\mu)}{3} \,\left( 24 + 24\,\xi  -
      24\,I_0(2\,{\sqrt{\xi }}) -
      2\,\xi \,I_0(2\,{\sqrt{\xi }}) \right. \nonumber\\&&
      \left. +
      2\,{\sqrt{\xi }}\,
       I_1(2\,{\sqrt{\xi }}) +
      {\xi }^{\frac{3}{2}}\,
       I_1(2\,{\sqrt{\xi }}) \right)\nonumber\\
       &&
\end{eqnarray}
$\xi$ being $|k\,w|$.

\end{document}